\documentclass[aps,prd,preprint]{revtex4}

\usepackage{enumerate}
\usepackage{amsmath,amssymb}
\usepackage{bm}
\usepackage{amscd}            
\usepackage{graphicx}

\begin{document}

\preprint{JLAB-THY-11-1312}

\title{Uncertainties in determining parton distributions
	at large $\bm{x}$}

\author{A.~Accardi}
\affiliation{Hampton University, Hampton, Virginia 23668 and\\
        Jefferson Lab, Newport News, Virginia 23606}
\author{W.~Melnitchouk}
\affiliation{Jefferson Lab, Newport News, Virginia 23606}
\author{J.~F.~Owens}
\affiliation{Florida State University, Tallahassee, Florida 32306}
\author{M.~E.~Christy, C.~E.~Keppel, L.~Zhu}
\affiliation{\mbox{Hampton University, Hampton, Virginia 23668}}
\author{J.~G.~Morf\'\i n}
\affiliation{Fermilab, Batavia, Illinois 60510\\}


\begin{abstract}
We critically examine uncertainties in parton distribution functions
(PDFs) at large $x$ arising from nuclear effects in deuterium $F_2$
structure function data.
Within a global PDF analysis, we assess the impact on the PDFs from
uncertainties in the deuteron wave function at short distances and
nucleon off-shell effects, from the use of relativistic kinematics,
and from the use of a less restrictive parametrization of the $d/u$
ratio. We find in particular that the $d$-quark and gluon PDFs vary
significantly with the choice of nuclear model.
We highlight the impact of these uncertainties on the determination of
the neutron structure functions, and on $W$ boson production and parton
luminosity at the Tevatron and the LHC.
Finally, we discuss prospects for new measurements sensitive to the
$d$-quark and gluon distributions but insensitive to nuclear corrections.
\end{abstract}

\maketitle

\section{Introduction}
\label{sec:intro}

Parton distribution functions (PDFs) are traditionally determined
{\it via} global fits to a wide variety of data for large momentum
transfer processes \cite{CTEQ6X,CT10,MSTW08,NNPDF,GJR,ABKM,HERAPDF}.
Typical data sets include deep inelastic scattering (DIS) of charged
leptons on proton and deuterium targets or neutrinos on heavy nuclear
targets, lepton pair production on proton and deuterium targets, and
the production of 
photons, vector bosons, or jets at large values of transverse momentum.
The use of deuterium targets in charged lepton DIS and heavy nuclear
targets for neutrino DIS brings in the added complication of having to
account for the modifications of the PDFs in a nuclear environment
compared to the PDFs of free nucleons.

One of the key goals of such global fits is the determination of the
flavor dependence of the extracted PDFs, {\it i.e.,} determining the   
PDFs for each parton flavor separately.  The classic method for
disentangling the $u$ and $d$ PDFs is to compare charged lepton DIS 
on proton and deuterium targets.
In order to concentrate on the extraction of the leading twist PDFs, 
cuts specifying minimum values of $Q^2 {\rm \ and \ } W^2$ are usually
employed, thereby reducing the contributions of higher twist terms and
target mass corrections.
The effect of these cuts is to restrict the range in $x$ to the region
$x \lesssim 0.7$.
In this region it is usually assumed that deuterium nuclear corrections
are a few per cent or less and are, therefore, negligible.
Accordingly, most global fits ignore deuterium nuclear corrections.

In our previous analysis \cite{CTEQ6X} the effects of systematically
reducing the cuts on $Q^2 {\rm \ and \ }W^2$ were studied.
This opened up the possibility of including data at larger values
of $x$ than those typically used in global analyses.
Target mass effects were included along with a parametrization of
higher twist terms.  It was found that the leading twist PDFs were
stable to variations in the techniques used to calculate the target
mass corrections, provided that a sufficiently flexible higher twist
parametrization was employed.
However, it was observed that the $d$ PDF was particularly sensitive to
the assumptions used to calculate the nuclear corrections in deuterium.
The purpose of the study reported here is to further investigate and
quantify the effects of nuclear corrections on PDFs in global fits which
make use of data obtained with deuterium targets, and to assess their
impact on the neutron structure function determination, as well as on
$W$ production and parton luminosities at the Tevatron and the Large
Hadron Collider (LHC).

In Section~\ref{sec:nuke} we present a detailed discussion of the
nuclear corrections which must be accounted for when using deuterium
targets, including those arising from the conventional nuclear Fermi
motion and binding, nucleon off-shell and relativistic effects, and
nuclear shadowing at small $x$.
Section~\ref{sec:data} contains a brief summary of the data used,
along with the fitting procedures, and a detailed discussion of the
flavor dependence of the various observables considered in this
study, with particular emphasis on the large-$x$ region.
Section~\ref{sec:fits} discusses the results of our fits, which
we refer to as ``CJ'' (CTEQ-Jefferson Lab) fits.
Future experimental possibilities to reduce the uncertainty in the
$d$-quark and gluon PDFs are outlined in Sec.~\ref{sec:discussion},
and our conclusions are summarized in Sec.~\ref{sec:conclusions}.

\section{Nuclear effects in deuterium}
\label{sec:nuke}

Because the deuteron is a weakly bound nucleus, most analyses have
assumed that it can be treated as a sum of a free proton and neutron.
On the other hand, it has long been known from experiments on nuclei
that a nontrivial $x$ dependence exists for ratios of nuclear to
deuterium structure functions.
While the ratio of the deuteron to isoscalar nucleon structure
functions has yet to be measured directly due to the absence of
free neutron targets, most theoretical studies suggest the presence
of non-negligible nuclear effects also in deuterium, although at
a level smaller than for heavy nuclei due to the deuteron's small
binding energy, $\varepsilon_d = -2.2$~MeV.
Particularly in the region $x \gtrsim 0.5$ the effects of nuclear
Fermi motion lead to a rapidly increasing ratio of deuteron to nucleon
structure functions, $R^{d/N} \equiv F_2^d/F_2^N$, which diverges as
$x \to 1$.
Any high-precision analysis of the large-$x$ region must therefore
account for the nuclear effects if data on deuterium (or other nuclei)
are used in the fit.

In this section we first review the standard nuclear corrections in the
deuteron, including the effects of nucleon Fermi motion and binding,
formulated within the framework of the weak binding approximation
(WBA) \cite{WBA,KP,KMK}.
For most kinematics of relevance to global PDF analyses a
nonrelativistic description of the deuteron is sufficient; however,
at large values of $x$ ($x \gtrsim 0.7$) the effects of relativity
become increasingly important.
We therefore explore the sensitivity of the nuclear corrections
to relativistic kinematics, as well as to the short-distance
nucleon--nucleon ($NN$) interaction parametrized through different
nonrelativistic and relativistic deuteron wave functions.
Finally, the dependence of the structure function of the bound
nucleon on its virtuality is explored within several models and 
phenomenological parametrizations and its impact on the $R^{d/N}$
ratio assessed.
Although not affecting this analysis significantly, for completeness
we also consider nuclear shadowing and meson exchange corrections in
the deuteron at small values of $x$.

\subsection{Standard nuclear corrections}
\label{ssec:conv}

The conventional approach to describing nuclear structure functions
in the intermediate- and large-$x$ regions is the nuclear impulse
approximation, in which the virtual photon scatters incoherently
from the individual nucleons bound in the nucleus \cite{West,Jaffe}.
Early attempts \cite{AW72,Poucher73,Bodek_smear,LP78,FS78} to
compute the nuclear corrections in deuterium, as well as some more
recent experimental analyses \cite{BCDMS1,BCDMS2,EMC1,EMC2,NMC1,NMC2},
utilized simple {\it ans\"atze} to account for the nuclear smearing,
some of which were rather {\it ad hoc} or suffered from various
shortcomings (for reviews see {\it e.g.} Refs.~\cite{Bickerstaff,GST}).
More recent efforts have focussed on deriving relations between the
deuteron and free nucleon structure functions more rigorously within
well-defined theoretical frameworks.

Within a covariant framework the nuclear structure function can
be written as a product of the virtual photon -- bound nucleon and
nucleon--deuteron scattering amplitudes \cite{Jaffe,MSTOFF},
which, for $x$ not too close to 1, can be systematically expanded
in powers of the bound nucleon momentum relative to the nucleon
mass, $\bm{p}/M$.
Keeping terms up to order $\bm{p}^2/M^2$, the nuclear structure
function can be written as a convolution of the bound nucleon
structure function and the distribution of nucleons in the nucleus
\cite{WBA}.

Because the struck nucleon is off its mass shell with virtuality  
$p^2 = p_0^2 - \bm{p}^2 < M^2$, its structure function can in
principle depend on $p^2$, in addition to $x$ and $Q^2$.
However, since the deuteron binding energy is only $\approx 0.1\%$
of its mass and the typical nucleon momentum in the deuteron is
$|\bm{p}| \sim 130$~MeV, the {\it average} nucleon virtuality
$\sqrt{p^2}$ will only be $\sim 2\%$ smaller than the free nucleon mass.
One can therefore approximate the bound nucleon structure function for
$x$ not too close to 1 by its on-shell value, in which case the deuteron
structure function can be written as a convolution of the free nucleon
(proton $p$ or neutron $n$) structure function and a momentum
distribution $f_{N/d}$ of nucleons in the deuteron,
\begin{eqnarray}
F_2^d(x,Q^2)
&\approx& F_2^{d\, \rm (conv)}(x,Q^2)\
=\ \sum_{N=p,n}
   \int_{y_{\rm min}}^{y_{\rm max}} dy\ f_{N/d}(y,\gamma)\ 
   F_2^N\left(\frac{x}{y},Q^2\right).
\label{eq:F2dconv}
\end{eqnarray}
The distribution $f_{N/d}$ (also called the ``smearing function'') is
in general a function of the deuteron's momentum fraction carried by
the struck nucleon,
  $y = (M_d/M) (p \cdot q/ p_d \cdot q)$,
where $q$ is the virtual photon momentum, and $p_d$ and $M_d$ are
the deuteron four-momentum and mass.
At finite $Q^2$, however, it also depends on the variable
  $\gamma = \sqrt{1 + 4 x^2 M^2/Q^2}$,
which in the deuteron rest frame can be thought of as the
virtual photon ``velocity'',  $\gamma = |\bm{q}|/q_0$.
In the Bjorken limit the function $f_{N/d}$ becomes independent
of $\gamma$, but for moderate $Q^2$ the dependence on $\gamma$
becomes important at $x \gtrsim 0.85$ \cite{KMK}.

The smearing function $f_{N/d}$ is computed from the deuteron wave
function $\Psi_d(\bm{p})$, and accounts for the effects of Fermi motion
and binding energy of the nucleons in the nucleus, as well as kinematic
$1/Q^2$ corrections; in the WBA the combined effects of these are given
by \cite{KP,KMK}
\begin{eqnarray}
f_{N/d}^{\rm (WBA)}(y,\gamma)
&=& \int{\frac{d^3\bm{p}}{(2\pi)^3}}
    \left( 1 + \frac{\gamma p_z}{M} \right)
    {\cal C}(y,\gamma)
    \left| \Psi_d(\bm{p}) \right|^2
    \delta\left( y-1-\frac{\varepsilon+\gamma p_z}{M} \right),
\label{eq:fWBA}
\end{eqnarray}
where $p_z$ is the longitudinal component of $\bm{p}$, and
the separation energy
$\varepsilon \equiv p_0 - M = M_d - E_p - M
             \approx \varepsilon_d - \bm{p}^2/2M$
by expanding the recoil nucleon energy $E_p = \sqrt{M^2 + \bm{p}^2}$
for small $\bm{p}$.
The finite-$Q^2$ correction factor is \cite{KP,KMK}
\begin{eqnarray}
{\cal C}(y,\gamma)
&=& \frac{1}{\gamma^2}
    \left[ 1 + \frac{(\gamma^2-1)}{y^2}
               \frac{(2 p^2 +  3 p_\perp^2)}{2M^2}
    \right],
\label{eq:CWBA}
\end{eqnarray}
where $p_\perp$ is the transverse nucleon momentum, and becomes unity
in the Bjorken limit, ${\cal C}(y,\gamma\to 1) \to 1$.
The resulting distribution function is sharply peaked around
$y \approx 1$, with the width determined by the amount of binding
(in the limit of zero binding, {\it i.e.}, for a free proton or
neutron, it would be a $\delta$-function at $y=1$).
At finite $Q^2$ (or $\gamma \neq 1$) the function becomes somewhat
broader, effectively giving rise to more smearing for larger $x$ or
smaller $Q^2$.
For nonrelativistic kinematics the variable $y$ is bounded by
  $y_{\rm min} = x (1 - 2 \varepsilon_d M/Q^2)$
and
  $y_{\rm max} = 1 + \gamma^2/2 + \varepsilon_d/M$,
so that its maximum allowed value, even in the Bjorken limit,
is approximately 1.5 instead of $M_d/M \approx 2$.

\begin{figure}
\includegraphics[width=10cm]{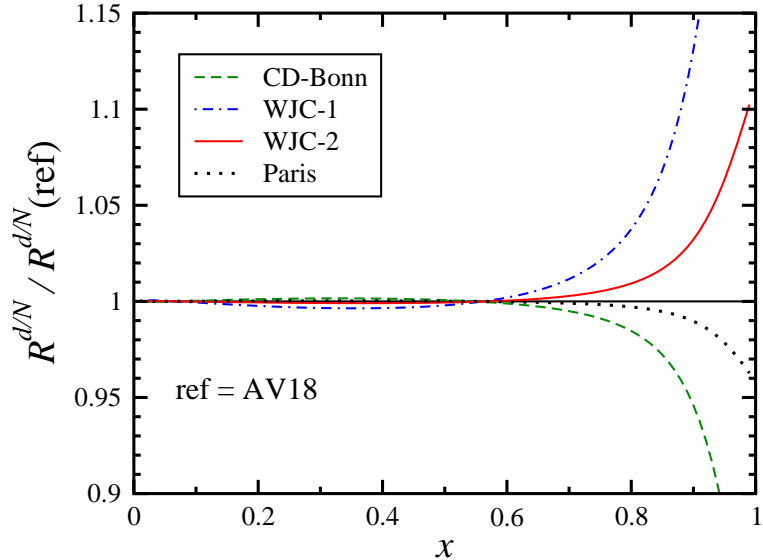}
\caption{Dependence of the ratio $R^{N/d} = F_2^d/F_2^N$ on the
	deuteron wave function, for the nonrelativistic CD-Bonn
	\cite{CD-Bonn} (dashed) and Paris \cite{Paris} (dotted)
	models, as well as the relativistic WJC-1 (dot-dashed)
	and WJC-2 \cite{WJC} (solid) models, relative to the
	ratio for the AV18 wave function \cite{AV18}.}
\label{fig:RR_wfn}
\end{figure}

The deuteron wave function $\Psi_d(\bm{p})$ contains the usual
nonrelativistic $S$- and $D$-states, and is normalized according to
$\int d^3\bm{p} \left| \Psi_d(\bm{p}) \right|^2 / (2\pi)^3 = 1$.
In the numerical calculations we use several nonrelativistic deuteron
wave functions, the high-precision AV18 \cite{AV18} and CD-Bonn
\cite{CD-Bonn} $NN$ potentials, as well as the older Paris potential
\cite{Paris} for comparison with earlier work.
As illustrated in Fig.~\ref{fig:RR_wfn}, where we show the computed
ratio $R^{d/N}$ relative to that for the AV18 wave function, the
differences between these are very small for $x \lesssim 0.7$.
At larger $x$ the differences become more visible, reflecting the
greater role played by the large-$y$ tail of the smearing function,
with the CD-Bonn wave function giving an approximately 10\% smaller
ratio for $x \sim 0.9$.
The result with the older Paris wave function agrees with the AV18
to within 1\% over this $x$ range.
The large-$y$ tail in turn reflects greater sensitivity to the
short-distance parts of the $NN$ interaction, where the validity
of the nonrelativistic approximation itself is more questionable.
For comparison we also show the ratios in Fig.~\ref{fig:RR_wfn}
computed with the relativistic WJC-1 and WJC-2 wave functions
\cite{WJC} (see Sec.~\ref{ssec:rel} below).  Because the relativistic
wave functions contain higher orders in $\bm{p}/M$, they exhibit
larger high-momentum tails, which leads to larger $R^{d/N}$ ratios
than for nonrelativistic wave functions.
Generally, the wave functions with the largest $D$-state probabilities
(7.4\% for WJC-1 {\it cf.} 4.8\% for CD-Bonn) give rise to the largest
distributions at large $y$, and hence the highest $R^{d/N}$ ratios at
large $x$.

\subsection{Relativistic effects}
\label{ssec:rel}

Since quarks at large momentum fractions $x$ are most likely to
originate in nucleons carrying a large momentum fraction $y$
themselves, the effects of relativity will be increasingly more
important as $x \to 1$.
A relativistic framework is therefore required to describe DIS
from the deuteron at very large $x$.
The effects of relativistic kinematics are generally straightforward
to implement, while the dynamical effects of relativity require model
dependent assumptions.

Note that for relativistic kinematics the maximum value for $y$ is
determined from $y_{\rm max} = M_d/M + x (M^2-p^2)/Q^2$; since the
nucleon virtuality $p^2$ is unbounded from below, at any finite $Q^2$
the fractional momentum $y$ is effectively unbounded from above.
In practice, though, taking $y_{\rm max} \approx M_d/M$, as would
be the case in the Bjorken limit, provides a relatively good
approximation to the exact integral for values of $x \lesssim 1$.

\begin{figure}
\includegraphics[width=10cm]{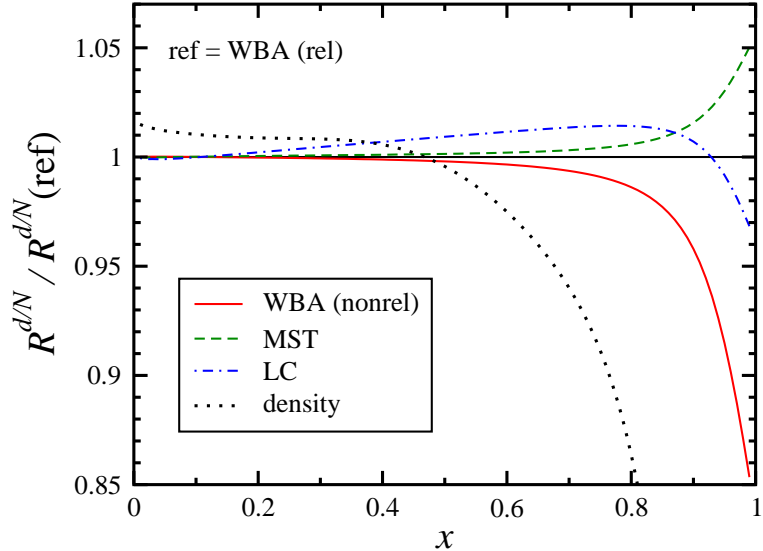}
\caption{Dependence of the ratio $R^{N/d} = F_2^d/F_2^N$ on the
	nuclear model and effects of relativity, relative to the
	ratio with the WBA model with the AV18 wave function
	and relativistic kinematics for $\gamma=1$.
	The nonrelativistic WBA model \cite{KP,KMK} (solid),
	the MST relativistic convolution \cite{MST} (dashed),
	and the light-cone model \cite{FS81} (dot-dashed)
	illustrate the nuclear model dependence, with the
	nuclear density extrapolation model \cite{density}
	(dotted) shown for reference.}
\label{fig:RR_mdl}
\end{figure}

The effects of relativistic kinematics on the WBA smearing function
(\ref{eq:fWBA}) are illustrated in Fig.~\ref{fig:RR_mdl}, where the
ratio $R^{d/N}$ for the smearing function with nonrelativistic
kinematics is plotted relative to the same ratio using the relativistic
expression for $y$, {\it i.e.}, not expanding the recoil energy $E_p$.
To isolate the kinematic effects on $R^{d/N}$, the same AV18 wave
function is used in both cases.  The relativistic effects are negligible
for $x \lesssim 0.6$, but grow to around 5\% at $x \approx 0.9$, before
increasing rapidly as $x \to 1$.

Taking the full spin and Lorentz structure of the off-shell nucleon
hadronic tensor into account, a formalism for DIS from bound nucleons
which incorporates both relativistic kinematics and dynamics was
developed in Ref.~\cite{MSTOFF}.
This analysis identified the conditions under which the nuclear
structure function could be expressed in the factorized convolution
form (\ref{eq:F2dconv}), but found that in general these are not
satisfied relativistically.
A follow-up study \cite{MST} (referred to as ``MST'') showed, however,
that one could isolate a ``relativistic convolution'' component from
the total $F_2^d$, together with additive corrections associated with
the relativistic $P$-state components of the deuteron wave function
and nucleon off-shell corrections.
Although the relativistic convolution component of $F_2^d$ is not unique
(only the total structure function is), a natural choice which respects
baryon number conservation in the deuteron (in the $\gamma \to 1$ limit)
is \cite{MST}
\begin{eqnarray}
f^{\rm (MST)}_{N/d}(y)
&=& \frac{ M_d^2}{4M}\
    \int \frac{d^3\bm{p}}{(2\pi)^3} \frac{y}{p_0}
    \left| \Psi_d(\bm{p}) \right|^2
    \theta(p_0)\,
    \delta\left( y - {p_0 + p_z \over M} \right).
\label{eq:fMST}
\end{eqnarray}
Expanding the integrand in Eq.~\eqref{eq:fMST} to order $\bm{p}^2/M^2$
one can show explicitly that $f^{\rm (MST)}_{N/d}$ reduces to the
nonrelativistic WBA smearing function in Eq.~(\ref{eq:fWBA}) in the
$\gamma \to 1$ limit.

In Eq.~(\ref{eq:fMST}) the relativistic deuteron wave function
$\Psi_d(\bm{p})$ formally contains, in addition to the $S$- and
$D$-state components, also the singlet and triplet $P$-state
contributions associated with negative nucleon energies.
To compute the relativistic smearing function we use the relativistic
deuteron wave functions from Ref.~\cite{WJC} with the WJC-1 model, which
has an admixture of pseudoscalar and pseudovector $\pi NN$ couplings,
and the WJC-2 model, which has only the pseudovector coupling.
The WJC-1 model generally has larger high-momentum components than
the WJC-2 model, as evidenced in its larger $D$-state probability,
and the pseudoscalar $\pi NN$ coupling leads to significantly larger
$P$-state wave functions than for the pure pseudovector model.
The harder WJC-1 wave function gives rise to a ratio $R^{d/N}$ which at
$x \gtrsim 0.6$ is larger than that for WJC-2, and as observed
in Fig.~\ref{fig:RR_wfn} the larger high-momentum tails in both the
relativistic wave functions yield larger contributions to $F_2^d$
(and $R^{d/N}$) at large $x$ than in the nonrelativistic models.

To assess the model dependence associated with the choice of smearing
function, in Fig.~\ref{fig:RR_mdl} we show the ratio $R^{d/N}$ computed
from the MST smearing function (\ref{eq:fMST}) relative to the WBA
model with relativistic kinematics (\ref{eq:fWBA}).  The differences
are $< 1\%$ for $x < 0.8$, rising to $\sim 5\%$ as $x \to 1$.
For comparison we also show the ratio computed using an early light-cone
prescription for nuclear DIS \cite{FS81}, which has been previously used
in some $F_2^d$ data analyses \cite{Whitlow}.
In this approach the bound nucleons are taken to be on their mass shells,
with the struck nucleon energy $p_0 = E_p$, but with energy not conserved
at vertices.  The light-cone smearing function (for $\gamma=1$) is given
by \cite{FS81}
\begin{eqnarray}
f^{\rm(LC)}_{N/d}(y)
&=& \int{\frac{d^3\bm{p}}{(2\pi)^3}}
    \left| \Psi_d^{\rm(LC)}(\bm{p}) \right|^2
    \delta\left( y - 1 - \frac{p_z}{E_p} \right), 
\label{eq:fLC}
\end{eqnarray}
which preserves baryon number and implies that the entire momentum
of the deuteron is carried by the nucleons.
Note that for consistency the light-cone wave function
$\Psi_d^{\rm(LC)}(\bm{p})$ 
in Eq.~(\ref{eq:fLC}) should be computed using light-front dynamics
(or in the infinite momentum frame), although in practice ordinary
rest-frame wave functions are used.
Because the light-cone prescription does not explicitly incorporate
nuclear mesons, which are responsible for nuclear binding, it
effectively corrects the deuteron structure function for the effects
of Fermi motion only.
This is reflected in the ratio $R^{d/N}$ in Fig.~\ref{fig:RR_mdl}
being larger at intermediate $x$ ($0.2 \lesssim x \lesssim 0.8$)
than in the models which do account for nuclear binding.
Nevertheless, the differences between $R^{d/N}$ computed with the
light-cone smearing function $f^{\rm (LC)}_{N/d}$ and the others
in Fig.~\ref{fig:RR_mdl} are less than 2--3\% for all $x$.
In practice, because of the inconsistency of mixing rest-frame wave
functions with the light-cone smearing function \eqref{eq:fLC},
as well as the need for finite-$Q^2$ corrections in the smearing
function (which are currently not available for the light-cone or MST
models), we shall restrict our analysis to the WBA smearing function.

Finally, we compare the microscopic calculations with a qualitative
model in which the ratio $R^{d/N}$ is extrapolated from data on the
ratio of heavy nuclei to deuterium structure functions assuming scaling
with nuclear density \cite{Yang}.
Using the empirical nuclear density for $^{56}$Fe and an {\it ansatz}
for the charge density of deuterium \cite{density,comment}, the ratio
is obtained from
	$R^{d/N} \approx 1 + (R^{{\rm Fe}/d} - 1)/4$
\cite{density}, which implies similar $x$ dependence of the nuclear
corrections in $^{56}$Fe and deuterium.
Consequently, $R^{d/N}$ in the density model has a significantly
larger depletion at $x \sim 0.6$ and a rise above unity which does
not set in until $x \sim 0.8$.
This is significantly higher than typically found in smearing
models, as reflected in the rapidly falling ratio $R^{d/N}$ in
Fig.~\ref{fig:RR_mdl} for the density model compared with the
other calculations \cite{Bodek91}.
The nuclear density scaling is also in disagreement with recent data
on light nuclei \cite{Seely} which do not support simple $A$-dependent
or density-dependent fits to the nuclear structure function ratios.
Because of this, as well as the difficulties in defining physically
meaningful nuclear densities for deuterium \cite{comment}, we do not
consider the density model sufficiently quantitative, especially at
large $x$, to be reliably employed in global PDF analyses.

\subsection{Off-shell corrections}
\label{ssec:off}

\begin{figure}
\includegraphics[width=10cm]{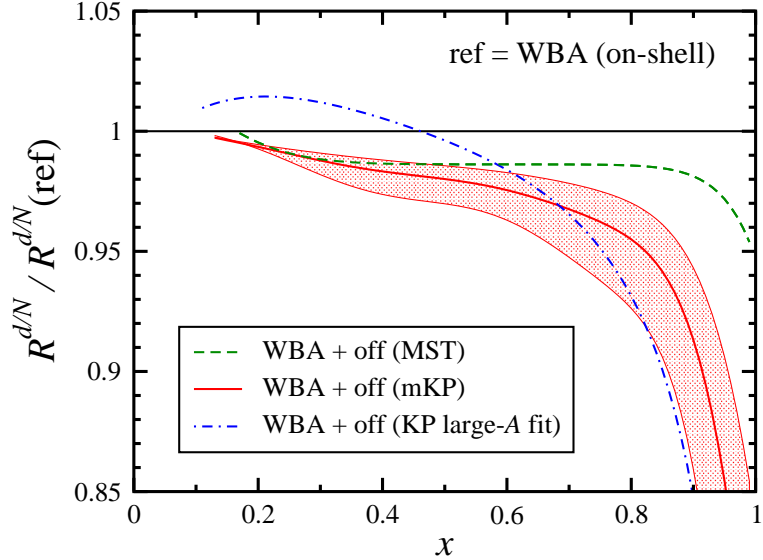}
\caption{Dependence of the ratio $R^{N/d} = F_2^d/F_2^N$ on the
	off-shell corrections to the nucleon structure function,
	relative to the on-shell WBA model.
	The range of off-shell behaviors includes the MST off-shell
	model \cite{MST} (dashed), the Kulagin-Petti model \cite{KP}
	modified for the case of deuterium (mKP, solid and shaded band),
	and the Kulagin-Petti phenomenological fit to large-$A$
	structure function data \cite{KP} (dot-dashed).}
\label{fig:RR_off}
\end{figure}

The discussion of nuclear effects above has been under the assumption
that the structure function of the bound nucleon in the convolution
formula (\ref{eq:F2dconv}) does not change inside the nucleus,
and can be approximated by the on-shell $F_2^N$.
With the inclusion of off-shell and relativistic corrections, it
was shown in Ref.~\cite{MST} that the convolution formula receives
nonfactorizable corrections associated with higher powers of the
bound nucleon momentum $\bm{p}/M$ and contributions proportional
to $p^2-M^2$,\
     $F_2^d = F_2^{d\, \rm (conv)} + \delta^{\rm (off)} F_2^d$.
The correction $\delta^{\rm (off)} F_2^d$ was estimated within
a simple quark--spectator model \cite{MST}, with the parameters
of the quark--nucleon correlation functions fitted to proton and
deuteron $F_2$ data.
The overall effect is a reduction of about 2\% of $F_2^d$ compared
to the on-shell (convolution) approximation, as illustrated in
Fig.~\ref{fig:RR_off}.
A simple parametrization of $\delta^{\rm (off)} F_2^d$ relative
to the total $F_2^d$ in the model of Ref.~\cite{MST} was given in
Ref.~\cite{CTEQ6X}.

Within the WBA model, inclusion of explicit off-shell dependence
in the bound nucleon structure function leads to a generalization
of Eq.~(\ref{eq:F2dconv}) in which $F_2^d$ is expressed as a
two-dimensional convolution in terms of $y$ and the nucleon virtuality
$p^2$ \cite{WBA,KP,KMK}.
For small values of $|p^2 - M^2|/M^2 \ll 1$, one can expand the
off-shell structure function in a Taylor series about $p^2=M^2$
\cite{KP},
\begin{eqnarray}
F_2^N(x,Q^2,p^2)
&=& F_2^N(x,Q^2)
    \left(  1 + \delta f_2(x,Q^2)\, \frac{p^2-M^2}{M^2}
    \right),
\label{eq:F2Noff}
\end{eqnarray}
where the coefficient of the $p^2-M^2$ term is given by the $p^2$
derivative of the structure function,
\begin{eqnarray}
\delta f_2(x,Q^2)
&\approx&
\left. \frac{ \partial \log F_2^N(x,Q^2,p^2)}{ \partial \log p^2 }
    \right|_{p^2=M^2}.
\label{eq:delf2}
\end{eqnarray}
Assuming that $\delta f_2$ is independent of the nucleus $A$,
a phenomenological fit to all available nuclear $F_2^A/F_2^d$
data, with $A$ ranging from $^4$He to $^{207}$Pb, found for the
nonrelativistic model \cite{KP}
\begin{eqnarray}
\delta f_2^{\rm (fit)}
&=& C_N (x-0.05) (x-x_0) (1+x_0-x),
\label{eq:delf2fit}
\end{eqnarray}
with the best-fit parameters $C_N = 8.1(3)$ and $x_0 = 0.448(5)$.
The zero crossings of $\delta f_2^{\rm fit}$ at $x=0.05$ and $x=x_0$
give rise to an enhanced $R^{d/N}$ ratio for $x \lesssim 0.45$, and
a sharp drop for $x \gtrsim 0.6$ compared with the on-shell result,
as Fig.~\ref{fig:RR_off} illustrates.
This reflects the general features of the data on the nuclear EMC ratios
$R^{A/d} \equiv F_2^A/F_2^d$, which the fitted off-shell correction
$\delta f_2^{\rm fit}$ mimics \cite{KP}.
Since one assumes the same shape for the off-shell correction in the
deuteron as in heavy nuclei, the $x$ dependence of $R^{d/N}$ in this
model is similar to that of $R^{A/d}$.
This feature is reminiscent of the nuclear density model \cite{density}
(see Sec.~\ref{ssec:rel}) which scales the magnitude of $R^{A/d}$
to the deuteron, but assumes the same $x$ dependence.
While this scaling may be reasonable for large-$A$ nuclei, we do not
believe that an extrapolation all the way to the deuteron is likely
to be reliable, especially at large $x$.
In addition, the phenomenological fit (\ref{eq:delf2fit}) does not
conserve the total baryon number in the nucleus.  Some cancellation
is found between the off-shell and nuclear shadowing corrections
\cite{KP}. However, since shadowing contains important higher twist
in addition to leading twist contributions, the cancellation cannot
be exact at all $Q^2$.

To avoid these problems we choose instead to use a microscopic model
proposed by Kulagin \& Petti \cite{KP} (see also Ref.~\cite{WBA}),
but modified to apply to the specific case of the deuteron in the
valence quark region at large $x$.  Here the valence quark distribution
can be expressed through a spectral representation as
\begin{eqnarray}
q_v(x,p^2)
&=& \int ds \int_{-\infty}^{k^2_{\rm max}} dk^2\, D_{q/N}(s,k^2,x,p^2),
\label{eq:qv}
\end{eqnarray}
where $k^2$ is the virtuality of the quark, with maximum value
$k^2_{\rm max} = x(p^2 - s/(1-x))$, and $s = (p-k)^2$ is the
mass of the spectator quark system.
The spectral function $D_{q/N}$ can be approximated by a factorized
form
\begin{eqnarray}
D_{q/N} &\approx& \delta(s-s_0)\, \Phi(k^2,\Lambda(p^2)),
\label{eq:DqN}
\end{eqnarray}  
where the function $\Phi$ describes the distribution of valence quarks
with virtuality $k^2$ in an off-shell nucleon with invariant mass
$p^2$, and for simplicity the spectator quark spectrum is assumed to be
given by a single mass $s_0$. The nucleon off-shell dependence
is parametrized through the $p^2$ dependence of the large-$k^2$
cut-off mass parameter $\Lambda(p^2)$ in $\Phi$.
From a fit to the free proton data one finds the value
$s_0 = 2.1$~GeV$^2$ \cite{KP}.
The scale $\Lambda$ can be related to the nucleon confinement radius,
$\Lambda \sim 1/R_N$, so that the dependence of $\Lambda$ on $p^2$
reflects a change in the size of the bound nucleon in a nuclear medium.
The off-shell function $\delta f_2$ can then be written as \cite{KP}
\begin{eqnarray}
\delta f_2
&=& c + \frac{\partial \log q_v }{ \partial x} h(x),
\label{eq:delf2model}
\end{eqnarray}
where
\begin{eqnarray}
h(x) &=& x(1-x) \frac{ (1-\lambda)(1-x) M^2 + \lambda s_0 }
		     { (1-x)^2 M^2 - s_0 }
\label{eq:gx}
\end{eqnarray}
and 
\begin{align}
\lambda =  \left. \partial \log\Lambda^2 / \partial \log p^2
		\right|_{p^2=M^2} \ .
\end{align}
In terms of the confinement radius $R_N$ the parameter
$\lambda$ can also be written as
	$\lambda = -2 (\delta R_N/R_N) (\delta p^2/M^2)$,
where
	$\delta p^2 = \int dy\, (p^2-M^2) f_{N/d}(y)$
is the average nucleon virtuality in the deuteron.

In the analysis of Ref.~\cite{KP}, $\lambda$ and $c$ were chosen
to reproduce the shape of the KP fit (\ref{eq:delf2fit}), finding
$\lambda=1.03$ --- a value that corresponds to an increase in the
confinement radius of $\delta R_N/R_N \sim 9\%$, as may be expected
for a heavy nucleus such as $^{56}$Fe.
Here we modify the model to the specific case of the deuteron,
for which $\delta R_N/R_N \approx 1.5$ -- 1.8\% \cite{CJRR}.
For $\delta p^2$ we take a range of smearing functions computed
with different wave functions (AV18, CD-Bonn, WJC-1, WJC-2),
as well as relativistic versus nonrelativistic kinematics,
resulting in the range $\delta p^2/M^2 \approx -3.6\%$ to $-6.5\%$.
This gives $\lambda$ in the range 0.46 -- 1.00, with a central
value of 0.65.  
Finally, the normalization coefficient $c$ is computed by requiring
that the off-shell corrections do not modify the valence quark number,
\begin{eqnarray}
\int_0^1 dx\, q_v(x)\, \delta f_2(x) &=& 0,
\label{eq:norm}
\end{eqnarray}
which leads to the constraint
\begin{eqnarray}
c &=& \int_0^1 dx\, \frac{\partial q_v(x)}{\partial x}\ h(x).
\label{eq:c}
\end{eqnarray}
The resulting ratio $R^{d/N}$ with the off-shell corrections in
this ``modified Kulagin-Petti'' (mKP) model is indicated in
Fig.~\ref{fig:RR_off} by the shaded band, which reflects the
uncertainty in the parameter $\lambda$.
For $x \lesssim 0.5$, the ratio with the mKP off-shell
correction is similar to that with the MST off-shell model,
ranging between 1\% and 3\%, but grows with increasing $x$.
At $x = 0.9$ the off-shell ratio is some 5\% -- 15\% smaller
than the on-shell ratio.
These uncertainties will propagate through the global analysis
and affect the precision to which PDFs at large $x$ can be
determined, as discussed in Sec.~\ref{sec:fits}.
Note that in principle the fits should be performed using the
same deuteron wave function to compute both the smearing function
and the off-shell correction (namely, for a specific value of
$\delta p^2/M^2$ rather than the above range).
However, given the uncertainty in $\delta R_N/R_N$, as well as
the assumptions inherent in the off-shell model itself, we take
the conservative approach of including the total off-shell
uncertainty in the fits for each smearing function.

\subsection{Nuclear shadowing}
\label{ssec:shad}

At intermediate and large values of $x$ the nuclear impulse
approximation, embodied in the convolution approximation of
Eq.~(\ref{eq:F2dconv}), provides a reasonable basis for
computing nuclear structure functions.
At small values of $x$, however, this approximation becomes
less reliable and the probability of scattering from more
than one nucleon in the nucleus becomes relevant.
For the case of the deuteron one can compute the shadowing
correction arising from the double scattering of the virtual
photon from both nucleons using the Glauber expansion in the
eikonal approximation (for a review see Ref.~\cite{SHAD_REV}).

In the target rest frame one can view nuclear DIS at small $x$
in terms of virtual photon fluctuations into $q\bar q$ pairs of
mass $m_{q\bar q}$, which then interact with the target.
The propagation length of the virtual $q\bar q$ state is
$\Delta l \sim 2 \nu / (Q^2 + m_{q\bar q}^2)$ with $\nu$ the
virtual photon energy.  If this exceeds the inter-nucleon
separation ($\sim 2$~fm), then the hadronic state can interact
with different nucleons as it passes through the nucleus.
For large $\nu \gg m_{q\bar q}$ the propagation length becomes
$\Delta l \sim 1/Mx \gtrsim 2$~fm, so that shadowing should start
to appear at $x \lesssim 0.1$.

At low $Q^2$ the spectrum of $q\bar q$ states can be well
approximated by vector mesons ($\rho^0, \omega, \phi$),  
while at high $Q^2$ the interaction is most efficiently
described through diffractive scattering of the $q\bar q$
pair from partons, parametrized by Pomeron exchange.
Such a two-phase model was adopted in 
Refs.~\cite{SHAD_KB,SHAD_MT93,SHAD_MT95,SHAD_PRW} to compute the
shadowing corrections in deuterium and heavy nuclei.
The vector meson dominance contribution vanishes at sufficiently
high $Q^2$, but in the $Q^2 \sim$~few~GeV$^2$ range it is in fact
responsible for the majority of the $Q^2$ variation \cite{SHAD_MT93}.

An additional double scattering contribution arises from the
interaction of the virtual photon with a mesons ($\pi, \rho, \ldots$)
exchanged between the nucleons in the deuteron \cite{SHAD_MT93,KU}.
This correction is mostly positive at small $x$ (``antishadowing''),
and partly cancels some of the shadowing arising from Pomeron exchange
and vector meson dominance \cite{SHAD_MT93}.

In our analysis we parametrize the total deuterium shadowing
correction from Ref.~\cite{SHAD_MT93} (similar values were obtained
in Refs.~\cite{SHAD_KB,SHAD_PRW,FGS}) as an additive contribution
to the deuteron structure function,
        $F_2^d \to F_2^d + \delta^{\rm (shad)} F_2^d$.   
The resulting correction ranges from about 1.5\% at $x = 10^{-2}$
to about 3\% for $x \leq 10^{-5}$.
The effect of such corrections on PDFs in the region of interest
of the current analysis is negligible.

\section{Data selection and fitting procedures}
\label{sec:data}

The data sets used in this analysis are the same ones chosen in the
recent CTEQ6X analysis \cite{CTEQ6X}.  Since this represents an
extension of the previous work, utilizing the same data sets
facilitates an easy comparison with the results presented there.
The data sets include charged lepton DIS on proton and deuteron
targets, lepton pair production with a proton beam on proton and
deuteron targets, $W$-lepton asymmetry and $W$ asymmetry data sets
from $\overline p p$ interactions, and $\gamma$ + jet and inclusive
jet data sets from the Tevatron.  
As in Ref.~\cite{CTEQ6X}, we adopt for our main fits the cuts
$Q^2 > 1.69$~GeV$^2$ and $W^2 > 3$~GeV$^2$.
Before proceeding with the discussion of the fits, it is useful to
briefly review the flavor differentiation at large $x$ provided by
the various types of data.

In the large-$x$ region the relevant quark PDFs are those for the
$u{\rm \ and \ }d$ quarks.  Consider first {\it charged lepton DIS}.
For large values of $x$ the structure functions are given, to lowest 
order in $\alpha_s$, by
\begin{equation}
F_2^p(x,Q^2) \approx \frac{x}{9}\left(4 u(x,Q^2) + d(x,Q^2)\right)
\end{equation}
and
\begin{equation}
F_2^N(x,Q^2) \approx \frac{5 x}{18}\left(u(x,Q^2) + d(x,Q^2)\right),
\end{equation}
where the superscript $N$ denotes the structure function of an
isoscalar nucleon target obtained from deuterium data by including
nuclear corrections.
If one had sufficient data of both types, then the separate
$u{\rm \ and \ }d$ PDFs could be disentangled.  The problem,
however, is that variations in the nuclear corrections cause
variations in the extracted $F_2^N$ which, in turn, causes
variations in the extracted $d$ PDF.  The $u$ PDF is relatively
better determined since, at large values of $x$, the $d/u$ ratio
is small and the $u$ PDF enters with a weight which is four times
larger than that for the $d$ PDF in $F_2^p$.  Thus, the nuclear
corrections primarily affect the $d$ PDF.

Next, consider {\it lepton pair production}.
Since the PDFs enter in a manner different than for charged lepton DIS,
one may hope to use these data to constrain the $d$ PDF.  Using lowest
order kinematics one has
\begin{equation}
M_{ll}^2 = x_1 x_2\, s \ \ \ {\rm and}\ \ \ x_F = x_1 - x_2,
\end{equation}
where $M_{ll}$ is the lepton pair invariant mass,
$x_F = 2 p_z/\sqrt{s}$ is the scaled longitudinal momentum
of the pair, and $x_1 {\rm \ and \ } x_2$ denote the beam and
target momentum fractions, respectively.
The region of interest for typical fixed target experiments is large
$x_1$ and small $x_2$, where the proton-proton cross section is
(neglecting the $Q^2$ dependence)
\begin{equation}
\sigma(pp)\ \propto\
\overline u(x_2)
\left( 4 u(x_1) + d(x_1)\, \frac{\overline d(x_2)}{\overline u(x_2)}
\right).
\end{equation}
At small values of $x_2 \ll 1$ one has
$\overline u(x_2) \approx \overline d(x_2)$ so that this cross section
is approximately given by the same linear combination of PDFs as enters
$F_2^p$, namely, $4 u(x_1) + d(x_1)$.
This situation is not helped significantly even if one considers lepton
pair production on a deuteron target since for small $x_2$ the nuclear
effects discussed previously are small and, using isospin invariance, 
the proton-neutron cross section is
\begin{equation}
\sigma(pn)\ \propto\
\overline d(x_2)
\left( 4 u(x_1) + d(x_1)\, \frac{\overline u(x_2)}{\overline d(x_2)}
\right).
\end{equation}
Thus, one is still sensitive to the same linear combination of PDFs
at large values of $x_1$, as probed by available data.

Next let us examine data for {\it jet production}.
The cross section for this process involves combinations of
quark--quark, quark--gluon, and gluon--gluon scattering subprocesses.  
The large-$x$ gluon PDF is constrained mostly by just such data,
assuming that the quark PDFs are fixed.
Therefore, if the $d$ PDF is varied it is possible for the gluon
PDF to compensate the change in the jet cross section by a similar
but opposite variation in its magnitude.
As a consequence, we can also expect the gluon PDF to be sensitive
to nuclear corrections, and anti-correlated to the $d$-quark PDF.

This leaves the data for the {\it $W$ asymmetry}, $A_W$,
and {\it $W$-lepton asymmetry}, $A_l$.
These asymmetries are sensitive to the $d/u$ ratio; however,
the data coverage extends at most to $W$ rapidities of
$y_W \approx 2.6$, which corresponds to $x \approx 0.6$.
The data for $A_l$ extend to lepton rapidities of $y_e \approx 3$.
This formally corresponds to $x \approx 0.8$, were it not for the
smearing induced by the lepton decay vertex, which reduces this to 
$x \approx 0.6$ (see Sec.~\ref{ssec:Wasy}).

Thus, one can see that to constrain the $d$ PDF in the large-$x$ region
using existing data one must use DIS data on deuterium, and there one is
faced with the necessity of addressing the issue of nuclear corrections.
In Sec.~\ref{sec:discussion} below, methods of reducing this dependence
on the nuclear corrections will be discussed.

\section{Fit results}
\label{sec:fits}

The global fits discussed here were performed using the same
fitting package and techniques as described in Ref.~\cite{CTEQ6X}.
We refer to these as the ``CJ'' (CTEQ-Jefferson Lab) PDF fits.
The nuclear effects in the deuteron were computed using the
smearing function (\ref{eq:fWBA}) with relativistic kinematics.
To quantify the systematic uncertainty on the PDFs due to the nuclear
corrections, we used the AV18, CD-Bonn, WJC-1 and WJC-2 deuteron wave
functions, with on-shell nucleons or off-shell corrections from the
range allowed in the mKP model, as discussed in Sec.~\ref{sec:nuke}.
We also considered a generalization of the $d$-quark parametrization at
the initial scale $Q_0^2=1.69$ GeV$^2$ that allows the $d/u$ ratio to
have an arbitrary value in the $x \to 1$ limit, as opposed to either
0 or infinity as with standard parametrizations used in global fits.

\subsection{Impact of nuclear corrections}

\begin{figure}
\includegraphics[width=10cm]{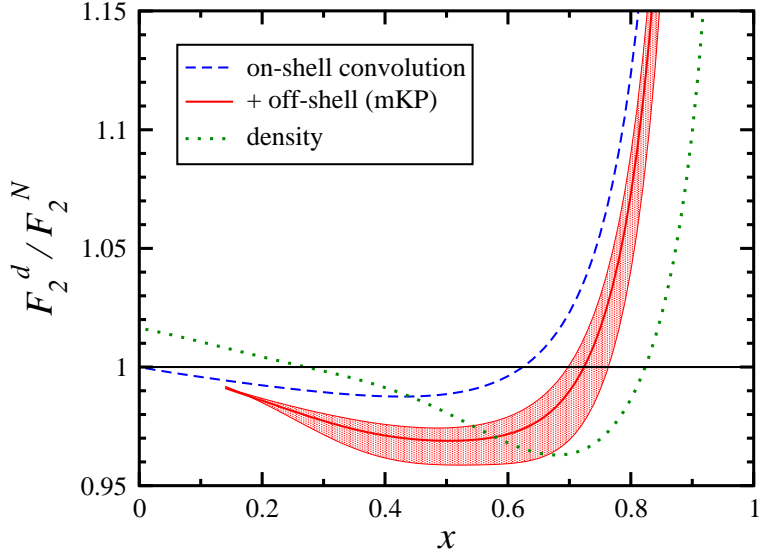}
\caption{Deuterium to isoscalar nucleon ratio $R^{d/N}=F_2^d/F_2^N$
	using the convolution approximation in Eq.~(\ref{eq:F2dconv}) 
	for a ``toy model'' input $F_2^N \sim (1-x)^3$, without
	off-shell corrections (dashed) and with the mKP off-shell
	corrections (\ref{eq:delf2model}) included (solid and
	shaded band).  For comparison the ratio with the nuclear
	density extrapolation model (dotted) is also shown.}
\label{fig:d2N}
\end{figure}

To understand the impact of the nuclear smearing and off-shell
corrections, it is useful to examine the ratio of $F_2$ on a deuterium 
target to that on an idealized isoscalar target $N=p+n$.
To simplify the comparison we take a ``toy model'' for the input
nucleon structure function, $F_2^N \sim (1-x)^3$, and compute the
deuteron structure function from the convolution approximation in
Eq.~(\ref{eq:F2dconv}) with the smearing function calculated using
the AV18 deuteron wave function.
The resulting ratio $R^{d/N}$ is illustrated in Fig.~\ref{fig:d2N}
(dashed curve), and compared with the ratio computed including nucleon
off-shell corrections from the mKP model (solid and shaded band),
and, for reference, also with the nuclear density extrapolation model
(dotted).
As anticipated from Fig.~\ref{fig:RR_off}, the effect of the off-shell
corrections is a further suppression of $R^{d/N}$ over most of the
range of $x$, with the ratio now rising above unity for $x \gtrsim 0.7$.
The dip in the $R^{d/N}$ ratio in the density model is more pronounced
than in the smearing models, even compared with the largest off-shell
corrections, and has a shape that essentially follows that of ratios
of structure functions of heavy nuclei, such as $^{56}$Fe, to deuterium.

Qualitatively, implementing the nuclear corrections is equivalent
to dividing the deuterium data by $R^{d/N}$, thereby yielding
results for an isoscalar target.  In actuality, the corrections
involve convolutions as well as realistic $F_2^N$ inputs instead
of the simple $\sim (1-x)^3$ form we use for illustration.
However, the impact of the nuclear corrections on fits of the
$d$-quark distribution can be understood from the following simple
argument, which does not depend on these details.
When $F_2^d/F_2^N$ is less than one, the removal of the nuclear
effects increases the data and therefore yields a larger $d$ PDF
than one would have had if the nuclear effects were ignored.
Conversely, if the ratio is greater than one, the fitted $d$ PDF
will be reduced relative to that extracted from uncorrected data.
From the ratio in Fig.~\ref{fig:d2N} one can see that there will be
an increase in the $d$ PDF for $x \lesssim 0.6-0.7$ (depending on
the size of the off-shell corrections), while the Fermi smearing
rise in the ratio leads to a decrease in the fitted $d$ PDF for
$x \gtrsim 0.6-0.7$.

\subsection{$\bm{d/u}$ ratio}

Since the $d$-quark PDF is most affected by variations in the nuclear
corrections, with the $u$ quark relatively unchanged, the results of
the fits are best summarized by examining the ratio of $d$ to $u$ PDFs.
This ratio is also of theoretical interest as its limiting value as
$x \to 1$ is a sensitive indicator of nonperturbative quark-gluon
dynamics in the nucleon \cite{nonpert,MT96}.
The effect of nuclear corrections on the gluon PDFs will be discussed
in Sec.~\ref{ssec:lumin}.

\begin{figure}
\includegraphics[width=\linewidth,bb=18 330 592 718,clip=true]
		{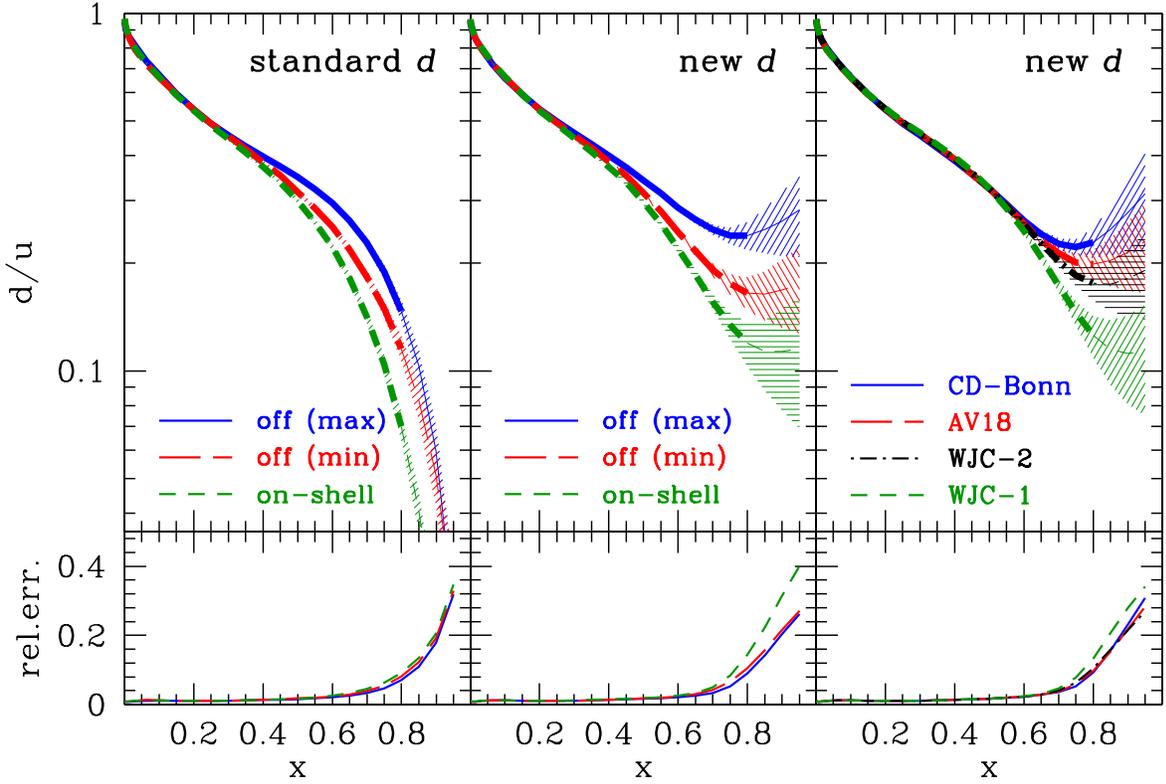} 
\caption{The $d/u$ ratio at $Q^2=10$ GeV$^2$ obtained with different
	choices of $d$-quark parametrization and nuclear corrections:
  ({\it left panel}) standard $d$-quark parametrization \cite{CTEQ6X}
	and AV18 deuteron wave function for several off-shell
	corrections models;
  ({\it middle panel}) modified $d$-quark parametrization
	(\ref{eq:dmod}) for the same nuclear corrections;
  ({\it right panel}) dependence on the deuteron wave function
	for a fixed off-shell correction (mKP).
	The shaded bands and bottom panels show the $\Delta\chi=1$
	PDF errors and the relative PDF errors on the $d/u$ ratio,
	respectively.  The thinner lines above $x \approx 0.8$
	denote extrapolations into unmeasured regions.}
\label{fig:d2uparams}
\end{figure}

In Fig.~\ref{fig:d2uparams} the $d/u$ ratio is shown for various
models of nuclear corrections. 
The left and middle panels show the variation induced by the
choice of nucleon off-shell corrections with a fixed (AV18)
deuteron wave function.
As the magnitude of the off-shell corrections increases,
the $d/u$ ratio correspondingly rises.
The left panel is obtained using a conventional parametrization
for the $d$ PDF which behaves as a power of $(1-x)$ as $x \to 1$ 
\cite{CTEQ6X,CT10}.
The middle panel shows the results of employing a modified form
of the $d$ PDF parametrization at the input scale $Q_0^2$,
\begin{equation}
d(x,Q_0^2)\ \longrightarrow\ d(x,Q_0^2)\ +\ a\, x^b\, u(x,Q_0^2),
\label{eq:dmod}
\end{equation}
with $a$ and $b$ fitted parameters (see also Refs.~\cite{Yang,Peng}).
With the conventional parametrizations of the $d {\rm \ and \ } u$
PDFs the $d/u$ ratio goes either to zero or infinity as $x \to 1$;
with the modified form the ratio $d/u \to a$ as $x \to 1$.
The fits, depending on the nuclear correction used, show a slight to
marked $\chi^2$ preference for a nonzero value of $a$, 
with its value increasing as the off-shell corrections become stronger. 
The error bands are larger for the case of the modified parametrization 
since the ratio is not required to go to zero at $x=1$, so one has a
wider range of possibilities for the ratio in the region beyond $x=0.8$.
Note, however, that the relative errors are essentially the same for
both parametrizations.

Variations in the choice of the deuteron wave function also affect
the extracted $d$ PDF, as shown in the right panel of
Fig.~\ref{fig:d2uparams}, where the modified $d$ parametrization
(\ref{eq:dmod}) is utilized.  The resulting spread is comparable in
magnitude to that observed for variations in the nucleon off-shell
correction models.

Note that the $d/u$ results above $x \approx 0.8$, marked by thinner
lines in Fig.~\ref{fig:d2uparams} for the central values, represent
extrapolations of the fitted PDFs and are not constrained directly by
data.  Correspondingly, the fit errors increase significantly for
$x \gtrsim 0.8$, especially for the PDFs using the less restrictive
$d$ parametrization (\ref{eq:dmod}), as noted above.  In all the fits 
the error bands
are calculated using the Hessian technique and correspond to
$\Delta \chi=1$, which was chosen for clarity of the presentation.
A larger choice of $\Delta \chi$ would have made it difficult to display
the effects of variations in the nuclear models.

\begin{figure}
\includegraphics[width=0.49\linewidth,bb=48 270 542 718,clip=true]
                {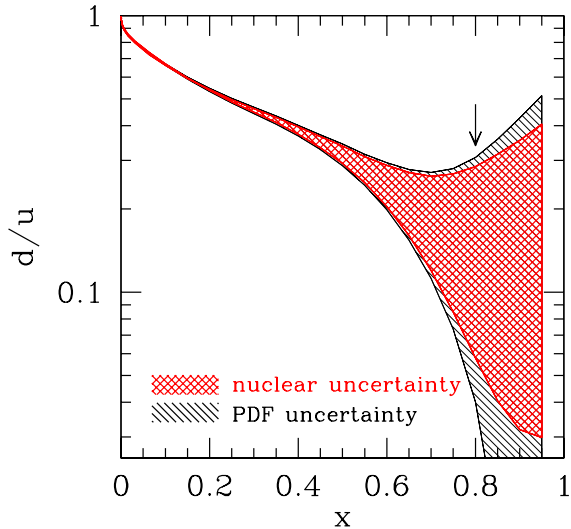}
\caption{Full range of variations of the $d/u$ ratio due to deuteron
	wave function and off-shell corrections (cross-hashed band)
	and PDF fit uncertainties (diagonal hashed).  The arrow
	indicates the extrapolation region at large $x$.}
\label{fig:du-total}
\end{figure}

Figure~\ref{fig:du-total} shows the total variation in the $d/u$ ratio
(cross-hashed band) for all the choices of deuteron wave function and
off-shell corrections considered in this analysis, using the modified
$d$-quark parametrization (\ref{eq:dmod}).
The highest $d$-quark PDF at large $x$ is obtained with the combination
of CD-Bonn wave function and the upper limit of the mKP off-shell
correction, while the lowest $d$ quark is obtained with the WJC-1
wave function and no off-shell corrections.
The outer band in Fig.~\ref{fig:du-total} (diagonal hashed) shows in
addition the uncertainty from the PDF fit.  As a reference fit for
later use we choose the PDFs extracted using the AV18 and mKP (central)
off-shell corrections, which produces a $d$-quark PDF approximately
midway between the upper and lower limits.
Above $x \approx 0.8$ the PDFs are not directly constrained by
data and represent extrapolations, as indicated by the arrow in
Fig.~\ref{fig:du-total}.

It is clear that the systematic uncertainty on the extracted $d$ PDF
due to the model dependence of the deuteron corrections is considerable.
One of the main interests of measuring the $d/u$ ratio at large $x$
is that its limit as  $x \to 1$ is sensitive to the nonperturbative
structure of the nucleon \cite{nonpert,MT96}.
However, given the large nuclear model uncertainty, essentially any
value of $d/u$ between 0 and $\approx 0.5$ is currently allowed,
which encompasses most values predicted by the nonperturbative models
\cite{nonpert}.

The nuclear model uncertainty on the $d/u$ ratio is several times
larger than the $\Delta\chi=1$ PDF uncertainty induced by the
experimental errors.  Other global fits, such as MSTW08 \cite{MSTW08}
or CT10 \cite{CT10}, utilize a different criterion to quote the PDF
experimental uncertainty, which can be approximately reproduced using
$\Delta\chi=7$.  In this case the PDF nuclear uncertainty would be
slightly smaller than the experimental uncertainty.  Overall, the
nuclear and experimental uncertainties are of the same order of
magnitude.  Progress in constraining the $d$-quark PDFs at large $x$
will therefore require either a better understanding of nuclear 
corrections to reduce the nuclear corrections systematics,
or the use of a much larger free-nucleon data set sensitive
to large-$x$ $d$ quarks than presently available.

\subsection{Fits of neutron structure functions}

\begin{figure}
\includegraphics[width=0.49\linewidth,bb=48 270 542 718,clip=true]
                {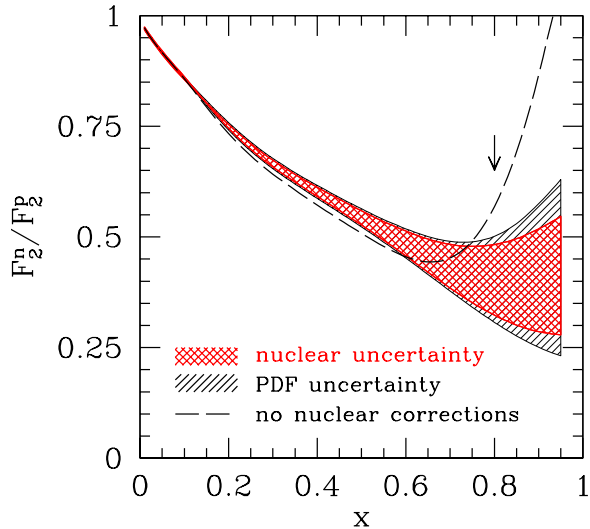} 
\caption{As in Fig.~\ref{fig:du-total}, but for the $F_2^n/F_2^p$
	ratio.  For comparison, the central value of a fit with no
	nuclear corrections applied is also shown (dashed).}
\label{fig:pn-total}
\end{figure}

The PDFs obtained from our global fit can be used to compute the
structure functions of a free neutron, providing an effective method
to extract these from inclusive deuteron DIS data, which additionally
benefits from information coming from a much wider range of processes.
In Fig.~\ref{fig:pn-total} we show the $F_2^n/F_2^p$ ratio calculated
including the full range of nuclear variations considered in our fits.
As in Fig.~\ref{fig:du-total}, the cross-hashed band illustrates
the variation due to the choices made for the deuteron wave function
and the off-shell corrections, while the hashed band shows in
addition the uncertainty coming from the PDF errors.
The arrow serves as a reminder that the bands at $x \gtrsim 0.8$
are an extrapolation unconstrained by data.
Note that as a consequence of the positivity of the $u$ and $d$
quark distributions, $F_2^n/F_2^p$ must be greater than 1/4 for
all values of $x$.
(The PDF error band can exceed that limit because it was obtained
with the Hessian technique, which assumes Gaussian errors also
close to the limits of the PDF functional space.)

For comparison we also show in Fig.~\ref{fig:pn-total} the
$F_2^n/F_2^p$ ratio extracted assuming no nuclear corrections in
deuterium.  Above $x \approx 0.75$ the neutron structure function
extracted under this assumption is significantly larger than the
range allowed by our fits.
(In fact, assuming $F_2^n=F_2^d-F_2^p$ implies a divergent
experimental $F_2^n/F_2^p$ ratio at $x=1$, where $F_2^p$ vanishes
but $F_2^d$ remains finite.)
The nuclear uncertainty in the $F_2^n/F_2^p$ ratio in
Fig.~\ref{fig:pn-total} is thus somewhat smaller at large $x$
than would be the case if we had considered the entire range of
possibilities for the nuclear effects, including the ``no nuclear
corrections'' {\it ansatz}.
Nonetheless, given the choice of the range of nuclear corrections
discussed in Sec.~\ref{sec:nuke}, the $F_2^n/F_2^p$ ratio at
$x \approx 1$ appears consistent with any value between 1/4 and
$\approx 0.7$, covering all nonperturbative model predictions
\cite{nonpert,MT96}.

A recent effort to extract the neutron structure function from
inclusive deuterium data \cite{Arrington,Rubin} follows a complementary
approach to that adopted here.  Firstly, phenomenological fits to proton
and deuteron structure functions are used to interpolate available data
to a fixed value of $Q^2$.  The data are then analyzed within the
convolution framework, using a range of nuclear corrections selected
from the literature \cite{Rubin}.
The resulting size of uncertainties on the $F_2^n/F_2^p$ ratio
in Ref.~\cite{Rubin} is smaller than but comparable to that in
Fig.~\ref{fig:pn-total}.  However, the central value was found to
decrease monotonically with $x$ and shows no sign of a plateau.
This is in contrast to our findings, which suggest a possible
plateau or even increase in $F_2^n/F_2^p$ for $x \gtrsim 0.7-0.8$.
An important difference between our analysis and that of
Refs.~\cite{Arrington,Rubin} is that our $F_2^n/F_2^p$ ratio is
determined from global fits of leading twist PDFs (which are
constrained to be positive definite), utilizing a wide array
of data, not restricted to hydrogen and deuterium DIS only.
The data analyzed in \cite{Arrington,Rubin}, on the other hand,
are at the structure function level, including possible higher twist
and other subleading $1/Q^2$ corrections, so that $F_2^n/F_2^p$ is
not constrained to lie above 1/4.

It is worth reiterating that there is a wealth of data in the
region up to $x = 0.98$ over a large range of photon virtualities,
$7 < Q^2 < 31$~GeV$^2$, which most previous global fits have
explicitly not utilized in order to avoid nonperturbative effects
such as nucleon resonances, target mass and higher twist corrections. 
Our fits include a sizable fraction of these data, delineated by the
cuts $Q^2 > 1.69$~GeV$^2$ and $W^2 > 3$~GeV$^2$, but still avoid most
of the resonance region data, and are limited to $x \approx 0.8$ as
indicated by the arrow in Fig.~\ref{fig:pn-total}.
These resonance region data are, however, included in the analysis
of Ref.~\cite{Arrington}, and may account for some differences in
the final results as compared to this work.   
Studies of the resonance region data using the standard nuclear
corrections demonstrate agreement between the DIS data sets and averages
over resonance region data \cite{Bosted,Malace,Accardi-HiX2010},
and future efforts will explore ways to incorporate the latter
within a global PDF framework.

\subsection{No-deuteron fits and the $\bm{W}$ asymmetry}
\label{ssec:Wasy}

\begin{figure}[t]
\includegraphics[scale=0.5,bb=18 144 470 718,clip=true]
                {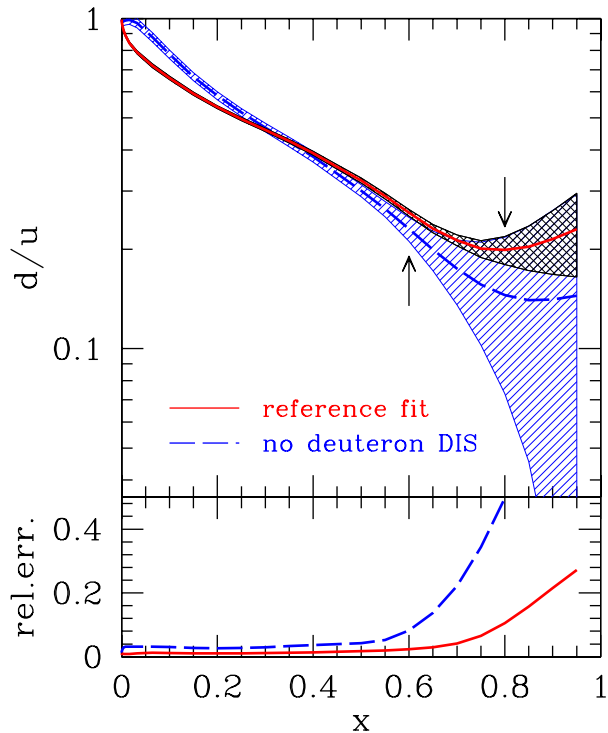}
\caption{$d/u$ ratios extracted from our reference fit, using the
	nucleon smearing function in the deuteron with the AV18
	wave function and the mKP (central) off-shell corrections.
	The solid curve utilizes all the data
	(with uncertainties given by the cross-hashed band);
	the dashed curve (with uncertainties given by the
	diagonal-hashed band) results from removing all the deuterium
	DIS data sets, avoiding the need for deuteron corrections.
	The PDF error bands correspond to $\Delta \chi=1$.
	The arrows indicate the extrapolation region of each fit
	($x > 0.6$ for the no-deuteron fit, and $x>0.8$ for the
	full fit).  The relative error of the two fits is shown in
	the bottom panel.}
\label{fig:noD}
\end{figure}

One of our motivations for including the deuterium data in the global
fit was to reduce the uncertainty on the $d$ PDF in the large-$x$ region.
The results presented thus far indicate that variations in the choices for
the nuclear models can be largely compensated by changes in the $d$ PDF.
The question thus naturally arises as to whether or not the addition of
the deuterium data has helped reduce the error on the $d$ PDF at all.
This can be examined by comparing in Fig.~\ref{fig:noD} the results of
two fits with and without the deuterium data; for the former, results
are shown for our reference fit using nucleon smearing in the deuteron
with the AV18 wave function and the mKP off-shell corrections.

Two things stand out in this figure.  First, for a specific choice of
the nuclear corrections, the $d$ uncertainty is significantly reduced,
as would be expected.  Thus, if a method could be devised which would
narrow the options for the choice of the nuclear models, then the
deuterium data would provide the information that is needed to
constrain the $d$-quark PDF.
On the contrary, in the absence of constraints on the nuclear models,
the systematic nuclear uncertainty is comparable to the PDF uncertainty
in the no-deuteron fit.
Secondly, the removal of the deuterium data results in a significant
increase in the $d/u$ ratio at {\it low} values of $x$, where the nuclear
smearing and off-shell corrections to deuteron data are negligible.
This was rather unexpected and suggests that the deuterium experiments
are pulling against some combination of other data in this region.
One might expect that in this region of low $x$ nuclear shadowing
corrections might be important; however, examination of various
shadowing models suggests a correction of only a percent or so,
insufficient to explain the shift in Fig.~\ref{fig:noD}.

\begin{figure}
\includegraphics[width=0.48\linewidth,bb=18 280 580 690,clip=true]
                {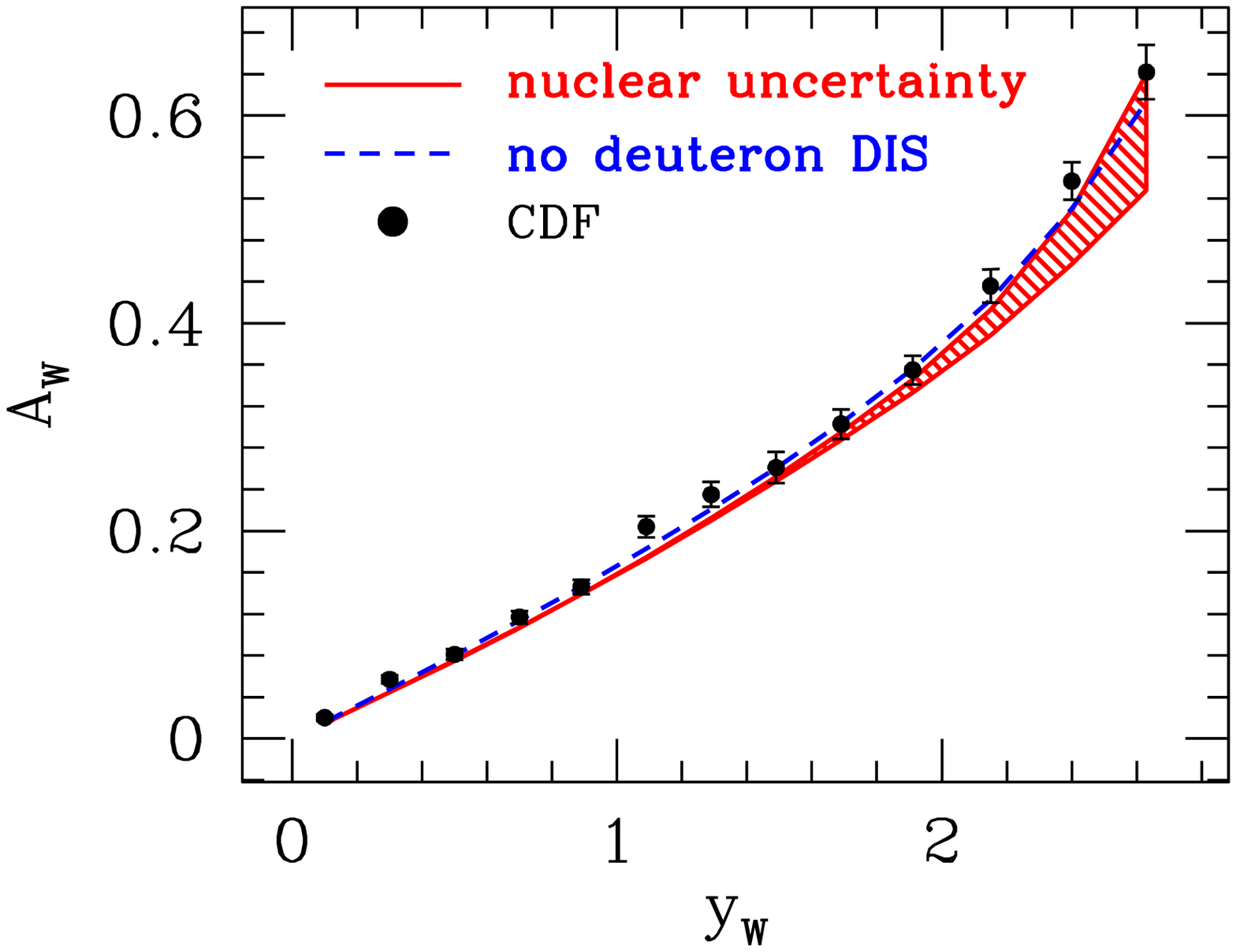}
\ \ \
\includegraphics[width=0.48\linewidth,bb=18 280 580 690,clip=true]
                {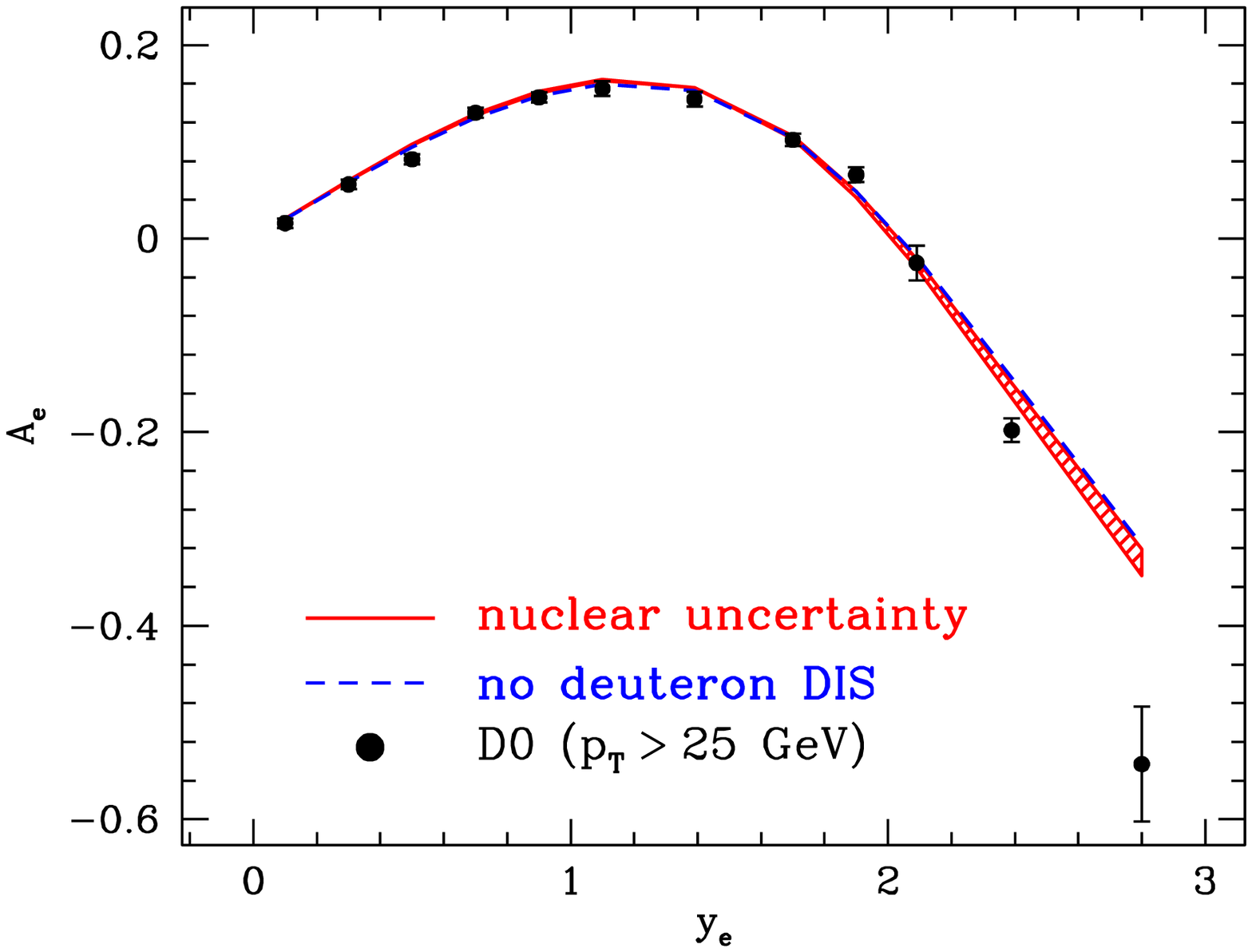}
\caption{
  {\it (Left)} The CDF $W$ asymmetry data compared to the
	calculation using fits with deuteron data and nuclear 
	corrections (shaded band) and the ``no deuteron DIS''
	fit (dashed curve).
  {\it (Right)} The D\O\ $W$ decay electron asymmetry data
	integrated over $p_T > 25$ GeV compared to the same fits.}
\label{fig:wasy}
\end{figure}

Inspection of the proton DIS data sets shows that the $\chi^2$
improvement between the reference and no-deuteron fits is slight
and is not confined to any specific region of $x$; this is true
also for the lepton pair and jet data.
However, the CDF $W$ asymmetry data, shown in Fig.~\ref{fig:wasy}
(left panel), suggest a marked decrease in $\chi^2$ of about 20
units at $W$ rapidity $y_W < 1.9$.
In contrast, for $y_W > 1.9$ the nuclear uncertainties become
non-negligible, making it more difficult to quantify the $\chi^2$
improvement of the no-deuteron fit, which is strongly dependent
on the reference nuclear model.   
These observations can be understood by noting that the $W$
asymmetry is approximately given by
\begin{equation}
  A_W \approx \frac{(d/u)(x_2) - (d/u)(x_1)}
	 	   {(d/u)(x_2) + (d/u)(x_1)},
\label{eq:approxAW}
\end{equation}
where $x_{1,2}=(M_{W^*}/\sqrt{s}) \exp(\pm y_W)$, and $M_{W^*}^2$
is the four-momentum squared of the virtual $W$ boson.
Equation~\eqref{eq:approxAW} is a simple leading-order result
in which the effects of the sea quarks are neglected and the
antiquark distributions in the antiproton have been set equal to
the quark distributions in the proton.  
It is clear that, at positive rapidity, increasing $d/u$ at small $x$
or decreasing $d/u$ at large $x$ results in an upward shift of $A_W$.

\begin{figure}
\includegraphics[scale=0.49,bb=18 360 470 700,clip=true]
                {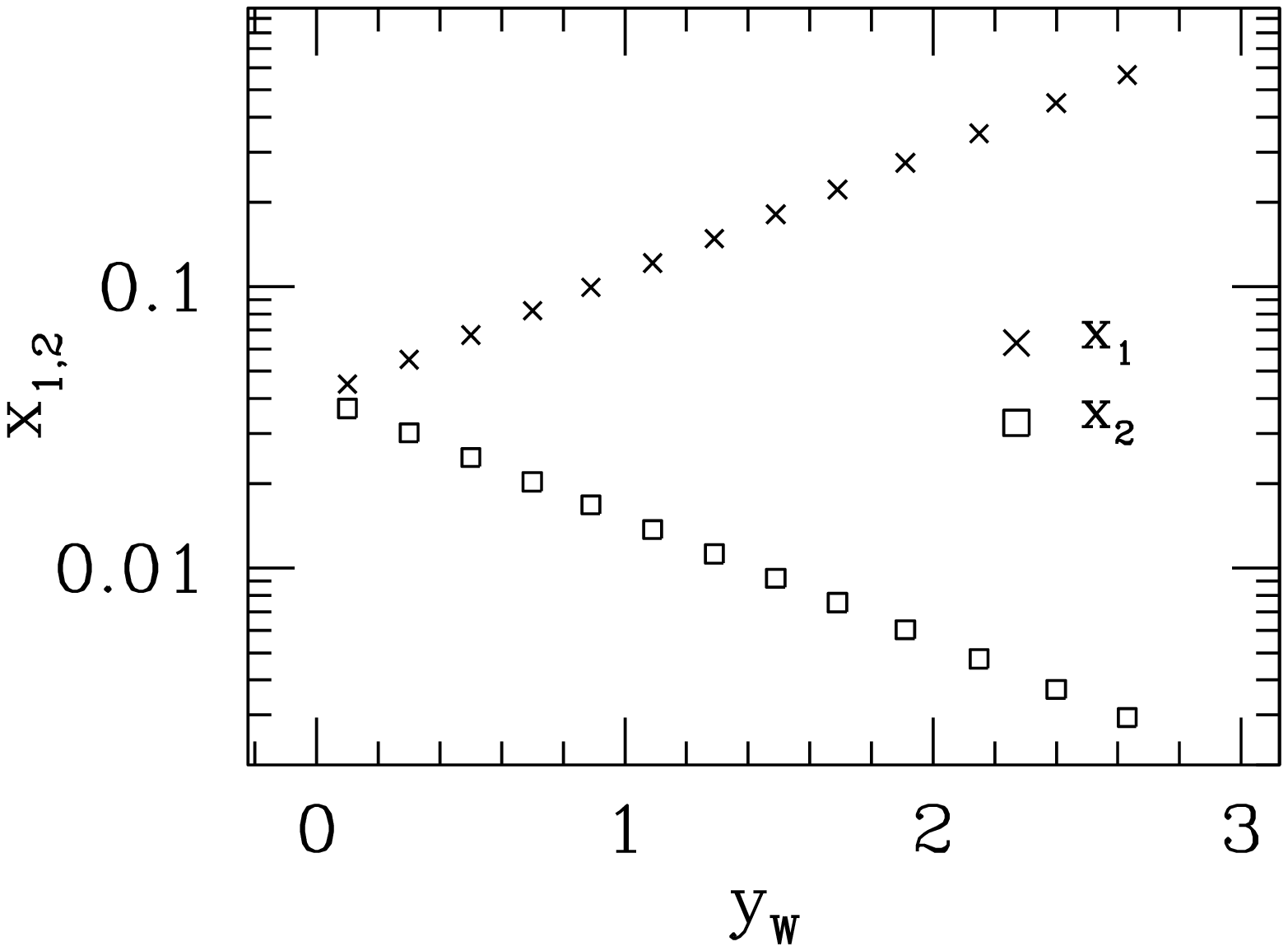}
\includegraphics[scale=0.49,bb=18 360 470 700,clip=true]
                {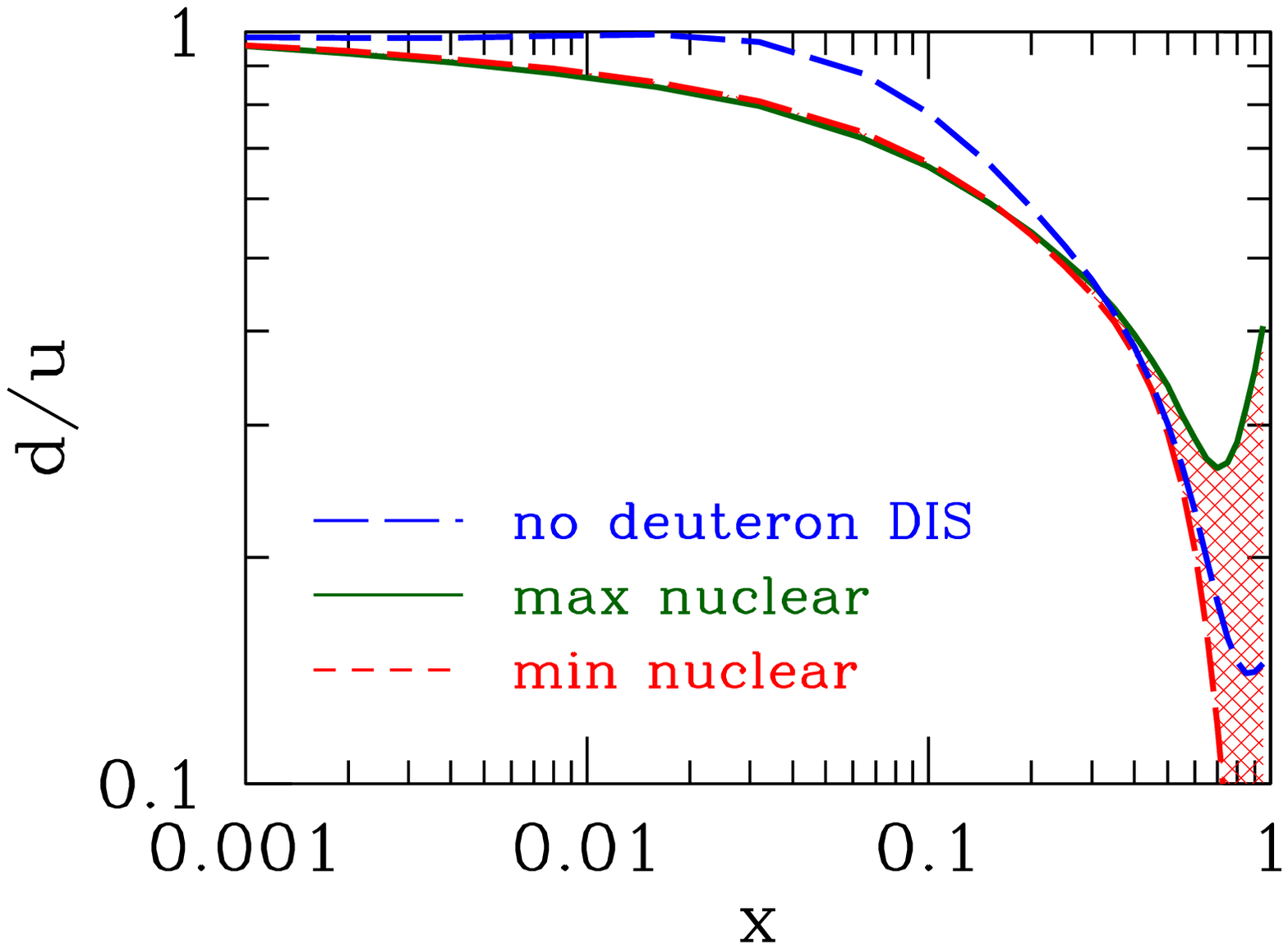}
\caption{
  {\it (Left)} Values of $x_{1,2}$ versus $W$ rapidity $y_W$ at
	CDF $W$ kinematics.
  {\it (Right)} $d/u$ ratios on a logarithmic $x$ scale for
	the ``no deuteron DIS'' fit (long dashed), and for
	maximum (solid) and minimum (short-dashed) nuclear
	corrections.}
\label{fig:du_logx}
\end{figure}

The interplay between $y_W$, $x_1$ and $x_2$ becomes clear from
Fig.~\ref{fig:du_logx}, which shows $x_{1,2}$ versus $y_W$ at
CDF kinematics (left panel), and the fitted the $d/u$ ratios on a
logarithmic $x$ scale (right panel) for the minimum and maximum
nuclear corrections, and the no-deuteron fit.
Data at $y_W > 2$ are sensitive to PDFs at $x_1 > 0.3$, where the
nuclear correction uncertainties become increasingly large, and at
$x_2 < 0.005$, where the difference between the full and no-deuteron
fits becomes small and the $d/u$ ratio is well constrained.
Large-$y_W$ data are thus mostly sensitive to large-$x$ partons,
and are affected by nuclear uncertainties.
Data at $y_W < 2$ are sensitive to intermediate values of
$x_{1,2} \in (0.01,0.1)$, precisely where the deuteron and
no-deuteron fits differ maximally.
It is clear that the improvement in $\chi^2$ at small $y_W$ is
entirely due to this region, where the deuteron DIS data and CDF
$W$ asymmetry data pull against each other.

The D\O\ collaboration data on the asymmetry $A_e$ of the electrons
originating from the $W$ decay, integrated over electron transverse
momenta $p_T>25$ GeV, is shown in the right panel of Fig.~\ref{fig:wasy}:
the small decrease in $\chi^2$ observed at small lepton rapidity $y_e$
is compensated by a small or large increase at large rapidity,
depending on the reference nuclear correction.
A similar behavior is observed in the D\O\ muon asymmetry data. 
Thus, in our fits, the lepton asymmetry data seem to pull only
very slightly against the deuteron target DIS data.
In this respect, one should note that due to the nature of the
vector-axial vector decay of the $W$, the lepton asymmetry is less
sensitive to the $d/u$ ratio than is the case for the $W$ asymmetry
itself. 

%
%

The $d/u$ ratio in the no-deuterium fit is significantly steepened in the 
low-$x$ region, as shown in Fig.~\ref{fig:noD}. This results in an increase 
in the $W$ asymmetry with a corresponding decrease in $\chi^2$. 
In an effort to understand the source of this behavior, 
the fitted PDFs were used to compare the theoretical predictions with the 
deuterium data which had been removed from the fit. 
The SLAC and BCDMS data start at $x=0.07$, the Jefferson Lab data at 
$x=0.40$, and the NMC deuterium/proton ratio data at $x=0.007$. The NMC ratio 
data have the most overlap with the relevant $x$ region for the low rapidity 
$W$ asymmetry and, therefore, they showed the biggest increase in $\chi^2$
when the reference and no-deuterium results were compared. Recently, Alekhin 
{\it et al.} \cite{Alekhin_NMC} have questioned the value of the ratio of the 
longitudinal and transverse cross sections, $R$, used to 
extract the structure functions from the NMC cross section data. However, as 
noted in Ref.~\cite{NNPDF_NMC}, the structure function ratio is relatively 
insensitive to $R$. One might also worry about the choice of parametrization 
for the PDFs. However, it is clear that the $W$ asymmetry data prefer a 
somewhat steeper $d/u$ ratio at small $x$ than do the deuterium data and no 
change in the form of the parametrization could alter the resulting tension 
since it occurs in a common range of $x$.

In summary, these plots and the $\chi^2$ inspection show that the
CDF $W$ asymmetry data and the D\O\ lepton asymmetry data seem
collectively to prefer a somewhat larger $d/u$ ratio than is favored
by the deuterium DIS data in the region below $x \lesssim 0.2$.
Moreover, as noted previously, at larger $x$ the lepton asymmetry
data prefer a somewhat larger ratio than the CDF $W$ asymmetry.
This unfortunately prevents them being used to constrain the nuclear
corrections, to which they are mildly sensitive.
The origin and possible resolutions of the tension between the $W$
and lepton asymmetries, as well as that between these asymmetries
and the deuteron data, is being actively discussed
\cite{MSTW08,CT10,Thorne,Ball,Guzzi}.

\subsection{Parton luminosities at hadron colliders}
\label{ssec:lumin}

\begin{figure}
\includegraphics[scale=0.7,bb=18 405 592 718,clip=true]{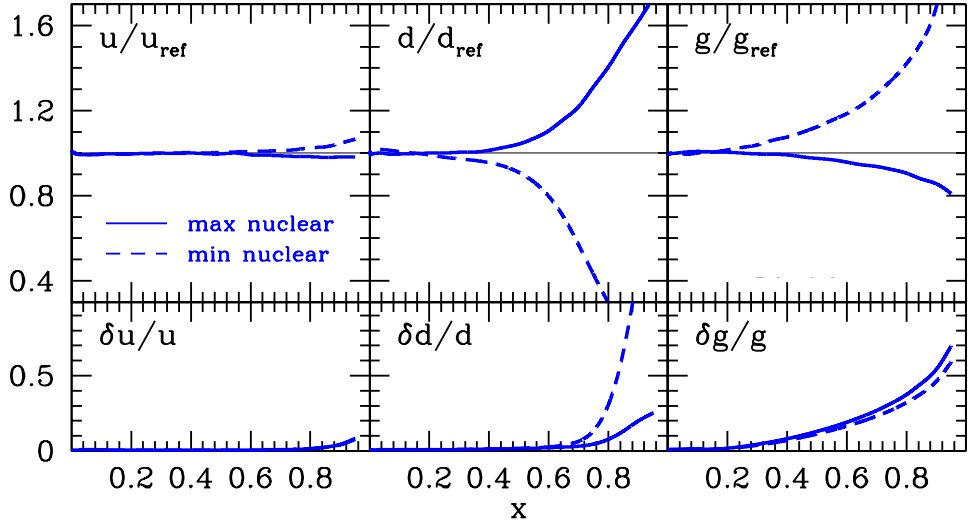}
\caption{The extremes of the variations of the $u$ {\it (left panel)},
	$d$ {\it (middle panel)}, and gluon {\it (right panel)} PDFs,
	relative to reference PDFs extracted using the smearing
	function with the AV18 deuteron wave function and mKP (central)
	off-shell corrections (\ref{eq:delf2model}).}
\label{fig:extremes}
\end{figure}

To discuss the effects of the nuclear model dependence on different
PDFs, the extremes of the $d$ PDF are shown in Fig.~\ref{fig:extremes}
divided by the reference fit PDF, along with the corresponding
$u$-quark and gluon PDFs.
Even though the $d$ PDF at large values of $x$ varies by as much as
$\approx 60\%$, the $u$ PDF shows little variation.
This is because the DIS and lepton pair data at large $x$ are for the
most part sensitive to the combination $4 u + d$, and the $d/u$ ratio
is on the order of 0.1 in this region.  Thus, the role played by the
$u$ PDF here is about 40 times more important that of the $d$ PDF.
Therefore, a small and anticorrelated shift in the $u$ PDF can
compensate for a large change in the $d$ PDF.

\begin{figure}
\includegraphics[scale=0.7,bb=18 475 592 718,clip=true]{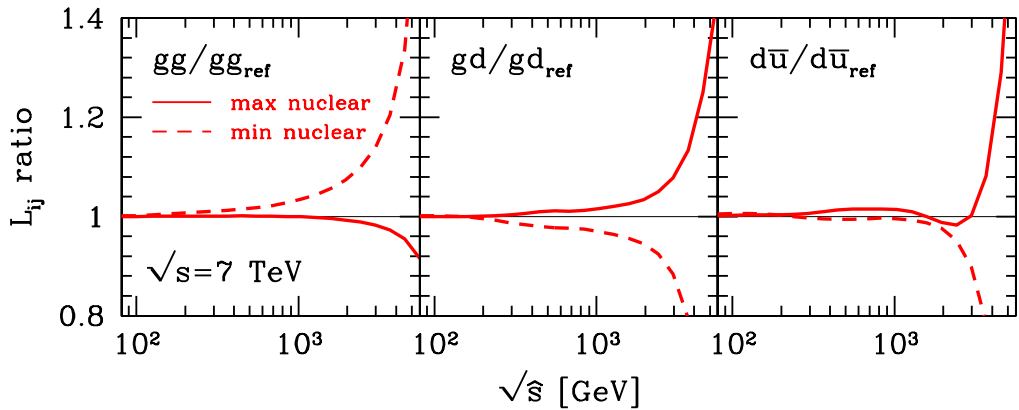}
\caption{Parton luminosities at $\sqrt{s}=7$~TeV at the LHC.
	Shown are the extremes of the variations due to deuterium
	nucleon corrections of the
	$gg$ {\it (left)},
	$gd$ {\it (middle)} and
	$d\bar u$ {\it (right)}
	luminosities relative to reference PDFs extracted using our 
	default nucleon smearing function in the deuteron with the
	mKP (central) off-shell corrections (\ref{eq:delf2model}).}
\label{fig:lum}
\end{figure}

\begin{figure}
\includegraphics[scale=0.7,bb=18 485 592 718,clip=true]{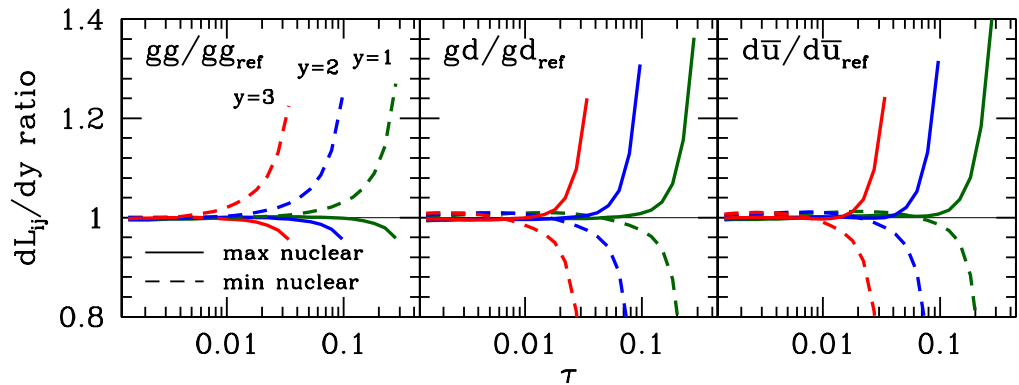}
\caption{Differential parton luminosities for
	$gg$ {\it (left)},
	$gd$ {\it (middle)} and
	$d\bar u$ {\it (right)}
	at fixed rapidity $y=1,2$ and 3, as a function of
	$\tau = \sqrt{\hat s / s}$, illustrating the variations
	due to the choice of deuterium nucleon corrections.}
\label{fig:lum_y}
\end{figure}

The situation is somewhat different for the case of high-$p_T$ jet
production.  This process receives contributions from $qq$, $qg$, 
and $gg$ scattering subprocesses with the $qq$ subprocesses
dominating at large values of $x$ for the target. 
Since the $u$ PDF is already well constrained, a variation in the
$d$ PDF is compensated by a significant and anti-correlated shift
in the gluon PDF. 
The sea quarks are largely unconstrained at large $x$, and follow
the behavior of the gluon, that generates them radiatively; the only
exception is the $\bar d$ quark, which seems to follow the $d$ quark
at $0.2 \lesssim x \lesssim 0.6$ and the gluon only at $x \gtrsim 0.6$.

The large nuclear uncertainty in the quark PDFs, and even more in the
gluon PDF, which shows a significant variation at values of $x$ as low
as 0.4, have potentially profound implications for future collider
experiments.  As an illustration, we consider the parton luminosities 
\begin{align}
  L_{ij} = \frac{1}{s (1+\delta_{ij})}
    \left[ \int_{\hat{s}/s}^1
	   \frac{dx}{x} f_i\left( x, \hat{s} \right) 
		        f_j\left( \frac{\hat{s}}{xs}, \hat{s} \right) 
	  + (i \leftrightarrow j)
    \right],
\end{align}
where $s$ ($\hat{s}$) is the hadronic (partonic) center of mass
energy squared, and $f_i(x,Q^2)$ is the PDF for a parton of flavor
$i$ (quark or gluon) at fractional momentum $x$ and scale $Q^2$.
In Fig.~\ref{fig:lum}, the parton luminosities are plotted 
at $\sqrt{s}=7$~TeV relevant to the current LHC runs.
The $gg$ luminosity controls the main channel for Higgs 
production, the $gd$ luminosity is relevant for jet production,
and the $d\bar u$ luminosity controls the ``standard candle''
cross section for $W^-$ production.
Figure~\ref{fig:lum}, shows that the theoretical uncertainty
induced by nuclear corrections extends to rather small scales
$\sqrt{\hat{s}}$, and grows quickly above 5--10\% as $\sqrt{\hat{s}}$
exceeds 1~TeV.

The luminosity ratios in Fig.~\ref{fig:lum} are relevant for 
the total cross section for producing an object with a mass equal to 
$\sqrt{\hat s}$. However, one can also investigate a specific region
of $x$ by considering the production of an object of the same mass,
but at a specified rapidity $y$ (not to be confused with the
momentum fraction $y$ carried by a nucleon in the deuteron,
as in Sec.~\ref{sec:nuke}). The differential parton luminosity
is then given by
\begin{align}
{dL_{ij} \over dy}
= \frac{1}{s (1+\delta_{ij})} 
  \big[ f_i\left( x_1, \hat{s} \right) 
	f_j\left( x_2, \hat{s} \right) 
        + (i \leftrightarrow j)
  \big],
\end{align}
where $x_{1,2} = \tau e^{\pm y}$, with $\tau = \sqrt{\hat s / s}$.
The differential luminosities, normalized to calculations using our
reference fit, are shown in Fig.~\ref{fig:lum_y} as a function of 
$\tau$ for three values of rapidity (since the ratios are largely 
independent of $Q^2$, these plots are also independent of $s$). 
The parton luminosities not plotted there show a qualitatively similar
behavior. 
Going to larger rapidities increases the sensitivity to the
large-$x$ region for a given mass.  Therefore, the nuclear corrections
come into play for smaller masses as the rapidity is increased. 
For example, for $W$ production at the Tevatron the nuclear
uncertainty becomes relevant at rapidity larger than 2, as already
observed in Sec.~\ref{ssec:Wasy}.

\section{Discussion}
\label{sec:discussion}

In the previous sections we have discussed in detail the fundamental
problem with using deuterium DIS data to constrain the $d$-quark PDF
at large $x$, namely, the deuterium data only constrain the
{\it combination} of the $d$ PDF {\it and} the model used to compute
the nuclear corrections.
Variations in the nuclear models are easily compensated by changes
in the $d$ distribution, while the $u$ distribution is already well
constrained by the hydrogen DIS data and other observables.
It is also interesting to note that the resulting variations in the $d$
PDF feed over to accompanying anticorrelated variations of the gluon PDF.
In order to disentangle the $d$ quark from the nuclear correction model,
one either needs additional experimental observables which do not rely
on nuclear targets, or experiments on nuclear targets for which it can
be arranged that the results are less sensitive to nuclear corrections.

\subsection{Constraints on large-$\bm x$ $\bm d$ quarks}
\label{sec:constrain-d}

One approach to directly constrain $d$-quark PDFs, avoiding nuclear
corrections altogether, would be to utilize data for $\nu p$ DIS,
which at large values of $x$ is dominated by the $d$ PDF.  
While such data exist \cite{WA21/25}, they were obtained using a
hydrogen bubble chamber and are statistically limited at large
values of $x$.  High statistics data for this process from new high
intensity neutrino beam lines, such as NUMI at Fermilab, would provide a
valuable constraint.
In addition, data on the ratio of $\nu p$ to $\bar{\nu} p$ cross
sections from a single experiment would provide direct sensitivity   
to $d/u$ at large $x$, while allowing for a partial cancellation in    
the uncertainties associated with beam flux.

A similar source of information on the $d$ PDF would be DIS data
for charged current $e^+ p$ interactions \cite{HERAe+}.
These data were taken as part of the HERA experimental program,
and will help constrain the $d$-quark PDF at small $x$, providing 
additional information concerning the observed tension between the
deuterium DIS data and the CDF $W$ asymmetry data.
At large $x$ they do not provide strong enough constraints due to   
limited statistics, even though a novel analysis is underway to extend
their reach to the largest possible $x$ \cite{Ingbir}.
Nevertheless, their inclusion in a full global fit should provide
additional constraints on the $d/u$ ratio.

Another observable sensitive to $d/u$ is the $Z$ rapidity
distribution, since the weak coupling of the $Z$ boson to the quarks 
makes it sensitive to a different combination of $u$ and $d$ quarks
from the Drell-Yan lepton pair data.
The cross section for $Z$-boson production involves a different linear
combination of parton luminosities than either $W$ production or lepton
pair production, thereby offering yet another constraint on the $d/u$
ratio.
We intend, in a future study, to investigate this observable
and include in our fits the available data, which currently have a
limited statistical significance at large $x$. 
A better experimental understanding of the $W$ asymmetry and
$W$-lepton asymmetry data is also needed, since at large rapidity
these data could help discriminate nuclear correction models.
However, since they pull the $d/u$ ratio in opposite directions,
they cannot yet be used to this purpose.

At lower energies, a number of experiments have been proposed at
Jefferson~Lab which seek to reduce or eliminate the need for the
nuclear corrections that limit current extractions of $F_2^n$,
and the next 5 years promises to significantly improve the
experimental situation with regards to the $d$ PDF at
$x \gtrsim 0.65$.
This will be provided through both flavor sensitive probes in
electron--proton scattering and electron scattering from
{\it effective} neutron targets.
Several of these experiments have already been approved for running
at Jefferson Lab after the upgrade of the accelerator to 12~GeV,
allowing for larger $Q^2$ and $x$ to be probed for $W^2$ outside
the nucleon resonance region.

One such method proposed for minimizing the nuclear uncertainties
in $F_2^n$ is the ``MARATHON'' \cite{MARATHON} experiment, which aims
to extract $F_2^n/F_2^p$ from a measurement of the $F_2$ structure
functions of tritium and $^3$He.  The use of these mirror nuclei allows
for the cancellation of nuclear effects to the $\approx 1\%$ level
\cite{A3}, as well as many of the systematic uncertainties.

A complementary method which has recently been successfully implemented
in the ``BONUS'' \cite{BONUS} experiment in Jefferson Lab Hall~B is
based on the spectator tagging technique.  Here, low momentum protons
tagged in the backward $\gamma d$ center-of-mass hemisphere in the
scattering of electrons from a deuteron target ensures that the
scattering took place off a nearly free neutron.
The effects of final state interactions of the neutron debris with
the spectator protons and nucleon off-shell corrections are
minimized by restricting the protons to momenta to below 100~MeV/c
and angles above $120^\circ$.
Utilizing the measured momentum vector of the proton and the virtual
photon, the initial momentum of the target neutron can be effectively
determined, allowing for the kinematic correction of Fermi motion
effects.  The results from the BONUS experiment with a maximum beam
energy of 5~GeV are limited to a maximum $x$ of about 0.65; however,
the power of the technique has been demonstrated and an extension
to $x \approx 0.85$ and higher $Q^2$ has been approved for running
at the 12~GeV energy \cite{BONUS12}.

Another novel, flavor-sensitive observable is provided by the
parity-violating DIS asymmetry (resulting from the interference of
$\gamma^*$ and $Z$-boson exchange) on a hydrogen target \cite{Souder},
which yields a new combination of $u$ and $d$ PDFs at large $x$,
and is free of any nuclear corrections.  A 6~GeV run has been completed
and subsequent runs are proposed for the 12~GeV era, over a range
extending to $x \approx 0.8$ \cite{SOLID}.

\subsection{Constraints on large-$\bm x$ gluons}
\label{sec:constrain-g}

As discussed in Sec.~\ref{sec:fits}, due to the interplay of deuteron
DIS data and large-$p_T$ jet data the gluon PDF is also very sensitive
to nuclear corrections.  Therefore these can also be constrained by
considering independent observables sensitive to large $x$ gluons.
Two such observables are the longitudinal DIS structure function $F_L$, 
and the charm structure function $F_2^c$ obtained by requiring at least 
one charmed particle in the final state of inclusive DIS.

The disadvantage of the $F_L$ structure function is that the gluon PDF 
typically decreases at large $x$ much faster than the $u$ and $d$ quarks, 
which dominate this region.  Therefore, in practice, the sensitivity of
$F_L$ to the gluons decreases fast as $x$ approaches 1.
The charm structure function is more directly related to the gluon
distribution, but variations in the heavy-quark scheme used in the
fit may in fact accommodate a number of possible gluon behaviors,
as has been observed in recent fits of structure functions at HERA
\cite{Cooper-Sarkar}.  It would nonetheless be worthwhile to include
$F_L$ and $F_2^c$ data in future fits to explore their power in
constraining large-$x$ gluons.

\subsection{Constraints on nuclear corrections}

The future free nucleon data discussed in
Section~\ref{sec:constrain-d} open the possibility of no-deuteron fits
with largely reduced $d$-quark PDF errors compared to that in
Fig.~\ref{fig:noD}, which uses currently available data sets. 
Having accurately determined the $d$-quark PDF at large $x$ 
one can then confront the validity of various nuclear correction
models directly. For instance, the free neutron $F_2$ structure
functions can be calculated from the fitted PDFs, and used in the
computation of the inclusive deuteron structure function. Any 
residual differences compared to data would then reflect directly the
choice of nuclear model.  
Likewise, Drell-Yan $p d$ data at {\it negative} rapidity, which
may be reached in collider experiments such as at RHIC, could be used
to further constrain the nuclear corrections in deuterium. 

Global QCD fits, however, provide an even more powerful method of
constraining the deuterium corrections, since they allow one to use a
more diverse set of data, and to take full advantage of the
statistical power of all the available data.  
For example, one could perform a
global fit additionally including data sensitive to gluons, such as
described in Section~\ref{sec:constrain-g}, and the deuteron DIS data
themselves. For each given choice of nuclear corrections the fit would
find PDFs, with very small PDF errors, that try to describe as well
as possible both the deuteron data and the free nucleon data.
A nuclear model that would, say, overcorrect the deuteron structure
function would then produce PDFs poorly describing either data set.
The phenomenologically viable nuclear corrections can then
be selected among those that minimize the tensions between these data.

Such tests of nuclear models of the deuteron within global PDF fits
will be complementary to direct experimental tests which will be
made possible by the future data, and are an important prerequisite
for an understanding of heavier nuclei.

\section{Conclusions}
\label{sec:conclusions}

Deep inelastic scattering and lepton pair data taken on proton targets
are generally sensitive to the combination $4 u + d$ at large values
of $x$.  An additional linear combination is required in order to
reliably separate the $u {\rm \ and \ }d$ PDFs.  Traditionally, this
role has been played by DIS data taken on a deuterium target.
In this analysis we have investigated the uncertainties induced in the
$d$ PDF by variations in the choices made for modeling the deuterium
nuclear corrections.  We have shown that reasonable choices for the
deuteron wave function and for modeling the off-shell effects result
in large uncertainties in the $d$ PDF as $x \to 1$.
As expected, the use of deuterium data reduces the uncertainty on the
$d$ PDF for any given set of choices for the nuclear models.  However,
the totality of the induced uncertainties for all of the models is large.

The $u$ PDF remains well determined despite the large uncertainties in
the $d$ PDF.  This is due to the fact that the large-$x$ DIS proton
data and the large $x_F$ lepton pair production data on both proton or
deuteron targets are sensitive to $4 u + d$, and at large values of $x$
the ratio $d/u \sim {\cal O}(0.1)$ so that the role played by the $d$ PDF
there is down by a factor of $\sim 40$ relative to that of the $u$ PDF.
Thus, a small change in the $u$ PDF can compensate a large change in the
$d$ PDF.  On the other hand, the $d$ and gluon PDFs are anticorrelated
by the high-$p_T$ jet data.  The large uncertainties of the $d$ PDF
lead to large uncertainties of the gluon PDF which extend down to
$x \approx 0.4$.  This has potentially profound implications for future
collider experiments.

This situation may be improved when data which are sensitive to the
$d/u$ ratio or gluons at large $x$, but which are {\it not} sensitive
to nuclear corrections, become available.  Several such experiments are
planned and should take data in the near future.  Inclusion of these
data in global QCD fits will help constrain the phenomenologically
viable range of nuclear correction models, thereby reducing the
currently large uncertainties in the $d$-quark and gluon distributions.

\begin{acknowledgments}

We thank F.~Gross, S.~Kulagin, S.~Malace, P.~Monaghan, V.~Radescu,
H.~Schellman and A.~Stadler for helpful discussions.  
This work has been supported by the DOE contract DE-AC05-06OR23177,
under which Jefferson Science Associates, LLC operates Jefferson Lab,
and NSF awards No.~0653508 and No.~1002644.  The work of J.F.O. is
supported in part by DOE contract number DE-FG02-97ER41022.

\end{acknowledgments}


\end{document}